\documentclass[11pt]{article}

\usepackage{amsmath}
\usepackage{amsfonts}
\usepackage{graphicx}

\newcommand{\mycaption}[1]{\caption{{\em #1}}}
\newcommand{\be}{\begin{eqnarray}}
\newcommand{\ee}{\end{eqnarray}}
\newcommand{\dd}{{\mathrm d}}

\newcommand{\half}{\frac{1}{2}}

\newcommand{\tG}{\tilde{G}}
\newcommand{\tC}{\tilde{C}}

\newcommand{\ZZ}{\mathbb{Z}}

\newcommand{\RR}{\mathbb{R}}

\newcommand{\DD}{{\mathbb{D}\, }}

\newcommand{\form}{\mathrm{form}}
\newcommand{\corr}{\mathrm{corr}}
\newcommand{\gauge}{\mathrm{gauge}}
\newcommand{\kin}{\mathrm{kinetic}}

\newcommand{\cs}{\mathrm{CS}}
\newcommand{\wz}{\mathrm{WZ}}

\newcommand{\so}{\mbox{\goth so}}

\newcommand{\myiota}{\iota} 
\newcommand{\myisharp}{{\myiota^*}} 
 
\newcommand{\newCC}{\Sigma}
\newcommand{\newC}{\sigma}
\newcommand{\newB}{b}
\newcommand{\newH}{h}
\newcommand{\cH}{\newH}
\newcommand{\newk}{\rho}
\newcommand{\newa}{\tau}
\newcommand{\newb}{\beta}
\newcommand{\newl}{\alpha}
\newcommand{\newe}{\gamma}
\newcommand{\newg}{\eta}

\newcommand{\newf}{\zeta}
\newcommand{\newq}{\xi}
\newcommand{\gravano}{{\cal I}}

\newcommand{\ft}[2]{{\textstyle{\frac{{\scriptstyle #1}}{\scriptstyle
#2}}}}

\newcommand{\sst}[1]{{\scriptscriptstyle #1}}

\newfont{\goth}{eufm10 scaled \magstep1}

\catcode`\@=11

\def\defop#1{\expandafter
\def\csname#1\endcsname{\mathop{\rm#1}\nolimits}}
\defop{tr}  \defop{Tr}  \defop{Ch}   \defop{ch}
\defop{Res} \defop{Det} \defop{SU}   \defop{SO} \defop{Diff}
\defop{str} \defop{Str} \defop{sdet} \defop{Sdet}
\defop{Map} \defop{Mat} \defop{diag} \defop{Sp} \defop{Spin}
\defop{Ad}  \defop{ad}  \defop{Tors} \defop{Lie}
\defop{Vol} \defop{vol} \defop{Aut}  \defop{Isom}
\defop{mod} \defop{Pexp}\defop{Out}  \defop{codim}
\defop{ind} \defop{supp}\defop{closure}

\def\norm#1{\mathopen\@@line#1\mathclose\@@line}
\def\slash{\@ifnextchar[{\@slash}{\@slash[\z@]}}
\def\@slash[#1]#2{
\setbox\z@\hbox{$#2$}\@tempdima\wd\z@\box\z@%
\@tempdimb#1 \advance\@tempdimb-\@tempdima \kern\@tempdimb
\hbox to\@tempdima{\hss\@makeslash\hss}}
\def\@makeslash{$/$}

\catcode`\@=12

\begin{document}

\begin{titlepage}
\begin{flushleft}
\hfill  Imperial/TP/01-02/25 \\
\hfill  IHES/P/02/57 \\
\hfill  {\tt hep-th/0212081} \\
\hfill  \\
\hfill  December, 2002 \\
\end{flushleft}
\vspace*{8mm}

\begin{center}

{\bf \LARGE Large Gauge Transformations in M-theory } \\
\vspace*{12mm}

{\large J.~Kalkkinen$^\mathrm{ab}$ and
       K.S.~Stelle$^\mathrm{a}$
} \\
\vspace*{4mm}

${}^\mathrm{a}$ {\em The Blackett Laboratory, Imperial College} \\
{\em Prince Consort Road, London SW7 2BZ, U.K.} \\

and \\  

${}^\mathrm{b}$ 
{\em Institut des Hautes \'Etudes Scientifiques, Le Bois-Marie} \\
{\em 35, Route de Chartres, Bures-sur-Yvette F-91440, France} \\

\vspace*{12mm} 

{\large\em Dedicated to the memory of Sonia Stanciu}
\vspace*{10mm} 

\end{center}

\begin{abstract}
We cast M-brane interactions including intersecting membranes and
five-branes in manifestly gauge invariant form using an arrangement of 
higher dimensional Dirac surfaces. We show
that the noncommutative gauge symmetry present in the
doubled M-theory formalism involving dual 3-form and 6-form
gauge fields is preserved in a form quantised over the integers. The proper
context for discussing large noncommutative gauge transformations is
relative cohomology, in which the 3-form transformation parameters
become exact when restricted to the five-brane worldvolume. 
We show how this structure yields the lattice of
M-theory charges and gives rise to the  conjectured 7D 
Hopf--Wess--Zumino term. 
\end{abstract}

\vfill

\noindent
{Email: {\tt jussi.kalkkinen@ihes.fr, k.stelle@ic.ac.uk} }

\end{titlepage}

\tableofcontents
\pagebreak


\section{Introduction}
\label{Introduction}

The classification of stable states in string and M-theory is
important not only for establishing the spectrum of the theory,
but also for understanding the physical equivalences of apparently
different configurations. In string theory, where we know how
branes interact through virtual open strings, the dynamics is well
in control, at least in the Ramond--Ramond sector 
\cite{Witten:1998cd,Bouwknegt:2000qt}. In M-theory,
this understanding is still largely missing for want of concrete
control of the higher unstable modes, as well as of models for coinciding
branes. Some aspects of the full interacting theory have been
discussed in terms of Matrix-theory, of holography and of anomaly
considerations. For instance, we know that the number of degrees
of freedom in a system of coinciding five-branes grows cubically as
a function of the number of branes
\cite{Harvey:1998bx,Bastianelli:2000hi}.

The M-theory spectrum contains, among other objects, membranes and
their magnetic duals, five-branes. In addition to these, there are
wave-like solutions and  gravitational instantons that bear
similarities to particles and six-branes, while boundaries of the
11D space can be interpreted as nine-branes
\cite{Horava:1996ma,Horava:1996qa}. Though these excitations also
reduce to D-branes in weakly coupled string theory, they are
inherently gravitational in nature. The worldvolumes of such
solitons also support solitons themselves: membranes may end on
five-brane worldvolumes, with the ends being described by
worldvolume self-dual strings
\cite{Strominger:1995ac,Townsend:1997em}. There are many more
examples.

The aim of this article is to find ways to describe this multitude
of branes and their intersections  simply in terms of geometry. It
appears that five-branes play a special r\^{o}le in this  and that
the geometrical description makes use of the {\em relative
cohomology} of the spacetime and five-brane worldvolume. The
elements of this cohomology come very close to being the
physically significant equivalence classes of brane
configurations. Related work on D2-branes in WZW has appeared in
\cite{Figueroa-O'Farrill:2001kz}. Other attempts in M-theory 
have been discussed {\it e.g.}~in 
Refs \cite{Freed:2000ta,deBoer:2001px}.

A more refined tool for analysing the geometry of M-branes is
provided by the gauge symmetry algebra that combines bulk global
(large) gauge transformations in bound M2--M5-brane systems with
bulk diffeomorphisms. This system is subject to chiral anomalies
on the five-brane worldvolume, requiring an extensive
anomaly-cancelling mechanism. The purely gauge subalgebra of this
relates to a ``doubled'' formalism \cite{Cremmer:1998px} of 11D
supergravity including both a 3-form $C$ and a 6-form $\tilde C$
gauge field; the essential commutator here is
$[\delta_3,\delta_3']=\delta_6$, where $\delta C=\Lambda_{\sst
3}$, $\delta \tC=\Lambda_{\sst 6}+\ft12 \Lambda_{\sst 3}\wedge C$
and for $\delta_6$ in the commutator one has $\Lambda_{\sst
6}=\Lambda_{\sst 3}'\wedge\Lambda_{\sst 3}$. The 3-form and 6-form
parameters of these transformations must be closed but need not be
exact, permitting thus also ``large'' gauge transformations,
depending on the cohomology of the underlying spacetime. In this
article, we shall refer to this algebra as the ``M-theory gauge
algebra.'' For earlier work on the doubled formalism {\it
cf.~e.g.}~Ref.~\cite{D'Auria:nx}. 

We shall show how the M-theory gauge algebra
survives in the global formulation of the theory, including the
large gauge transformations, provided that one refines the notion
of cohomological nontriviality to that of nontriviality in relative
cohomology: $\Lambda_3$ must reduce to an exact form when
restricted to five-brane worldvolumes.

There are several special features of M-theory five-branes that we
shall have to contend with. The first observation is that, although
the static five-brane solutions are non-singular, fluctuations
around them become singular owing to concentration of waves at the
horizon. Accordingly, one needs to introduce a $d=6$ worldvolume
action for the five-brane
\cite{Witten:1997hc,Pasti:1997gx,Aganagic:1997zq,Cederwall:1997gg}, ({\it
cf.}~also \cite{Bastianelli:hi});
this acts as a source for the external 11D gauge fields and
gravity \cite{Cremmer:1978km}. The introduction of such a
delta-function source will require careful regularisation at the
brane surface, however.

Another distinctive feature is the way in which the cancellation of anomalies
under chiral transformations of worldvolume determinants takes place.
Though the bulk theory is known to contain $C R^4$-terms that give rise under
diffeomorphisms to anomaly inflow onto the five-brane worldvolume, this is not
quite sufficient to cancel all of the chiral anomalies, since an anomalous term
proportional to the second Pontryagin class of the normal bundle remains
\cite{Witten:1997hc}. Cancelling this part of the anomaly seems to involve
understanding the details of the geometry of the $C$-field \cite{Freed:1998tg}. 
This would require a detailed description of the behaviour of the 
$C$-field solution under diffeomorphisms, which we do not include in this article. 

The plan of the paper is as follows: we describe the dynamics of
interacting M-branes perturbatively  in Section
\ref{Perturbative}. In Section \ref{Strings} we formulate the
model in such a way that invariance under large gauge symmetries
becomes manifest. The independence of this description from the
various choices of Dirac surfaces that we have to make is shown in
Section \ref{Consistency}. This provides a new derivation for the
known relationships between the brane tensions. Finally, Section 
\ref{Algebra} contains a short description of the gauge and 
diffeomorphism algebras. We have tried to make
the paper self-contained both physically and mathematically; 
many of the mathematical tools that we use are explained in the appendices.

\section{Perturbative supergravity}
\label{Perturbative}

The bosonic part of the 11D supergravity action
\cite{Cremmer:1978km} involves the metric and a 3-form $C$. The
form-field part can be written as
\be
S_\form &=&  ~\frac{1}{\kappa} \int_X  \frac{1}{2\newg}~ G \wedge
* G - \frac{1}{6} ~C \wedge G \wedge G
 \label{bulk} ~,
\ee
where $G \equiv \dd C$  is the field-strength and $X$ is the 11D spacetime. We will later
consider gravitational corrections to this action. The 3-form field can
be naturally coupled to membranes while its electromagnetic dual
6-form $\tC$, to be more properly defined later, couples to five-branes.
The five-brane supports a $d=6$ chiral $N=2$ tensor
theory on its worldvolume. This theory involves  an antisymmetric  
rank-two tensor
$\newB$, with (anti-)self-dual\footnote{We are at liberty to consider
either self-dual or anti-self-dual fields $\newH$ just by choosing the
sign of $\newl/\newe$ accordingly. We can therefore  present the
calculation without loss of generality for the self-dual choice of
fields.} field strength $\newH$, and five real scalars. This means
that we should couple the bulk action to sources of the form
\be  
S_{\cs} &=&
 \int_{W_6}  \frac{\newl}{2} \newH
\wedge *\newH + \newb~ \myiota^*\tC + \newe~ \newH \wedge \myiota^*C   \nonumber \\
& & \qquad + \newk \int_{W_2} \newB + \newa \int_{W_3} C  \label{cs} ~.
\ee
We will fix the coefficients in  units of the Newton
constant $\kappa$ presently, by requiring gauge invariance. Gauge
invariance will also impose geometrical constraints on the
currents, {\it cf.}~App.~\ref{Currents}, which appear here as volume forms
for integration domains.

\begin{figure}[ht]
\begin{center}
\includegraphics{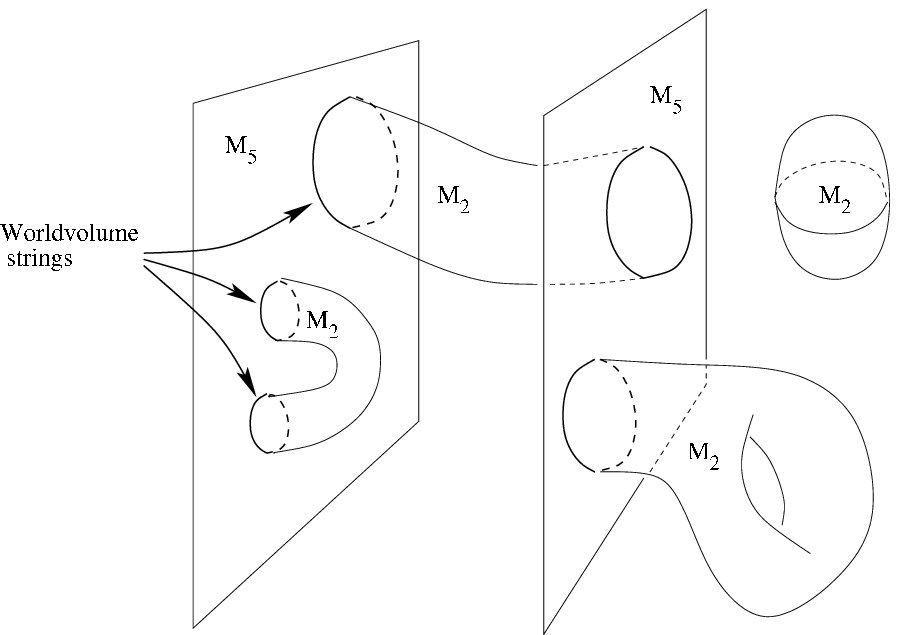}
\mycaption{Some M-theory solitons.} 
\label{fig0}
\end{center}
\end{figure}

Apart from the fermions, we will not pay particular attention here to a number of
other worldvolume interactions. 
For instance the kinetic terms for bulk gravity and the worldvolume 
scalars follow from
\be  
S_{\kin} &=&
\frac{1}{\kappa \eta^3} \int_X {\cal R} *1 + 
T_6 \int_{W_6} \dd^6 \sigma^i \sqrt{|\det G_{ij}|}  \nonumber \\ 
& & + 
T_3 \int_{W_3} \dd^3 \sigma^\mu \sqrt{|\det G_{\mu\nu}|} ~,
\ee
where $G_{ij},G_{\mu\nu}$ are suitable pull-backs \cite{Witten:1997hc,Pasti:1997gx,Aganagic:1997zq,Cederwall:1997gg} 
of the bulk metric and $T_3$,$T_6$ are brane tensions. There are many more. 
Neither will we include explicit dependence on worldvolume scalars, 
as these, as well as other twisting data such as the covariant derivative on the five-brane 
normal bundle, are implicit in the inclusions $\myiota$ at the classical level. 
Furthermore, Wess--Zumino terms built out of these scalars and covariant derivatives
would  transform under transverse diffeomorphisms only, and are therefore not
directly related to the gauge symmetries considered here; 
they play a r\^{o}le in anomaly cancellation, however. 

What singles out the retained terms in $S_{\form} +S_{\cs}$,
though, is the fact that these are the only gauge non-invariant
Chern--Simons type terms that we can write down, {\it i.e.~}terms
of the form $\dd^{-1} \omega$, where $\omega$ is some closed form.
This means that these are the only terms that will transform
nontrivially under gauge transformations, while all others remain inert by
definition.
 
In addition to these worldvolume terms it turns out to be necessary 
to consider further Wess--Zumino type contributions. These terms
have no apparent {\em raison d'\^{e}tre} from the point of view of
the physical 11D spectrum, but will turn out to be crucial in defining a
manifestly gauge invariant action using Dirac surfaces:
\be
S_{\wz}  &=&
 \newf \int_{W_7} C \wedge G  + \frac{\newq}{2} \int_{W_8} G \wedge G
 \label{wz}~.
\ee
In particular, these terms will prove to be essential for obtaining the 
correct charge lattice. Here, too, we leave the integration domains as well as the
coefficients as yet undetermined, which will allow us to retain the option of
setting them to zero later if necessary. The total action is then the sum
\be
S = S_\kin + S_\form + S_{\cs} + S_{\wz} \equiv  S_\kin + S_\gauge \ . 
\ee  
We will restrict our attention to the part of this relevant to gauge symmetry,
$S_\gauge$.  The constraints on brane
tensions will guarantee that the Wess--Zumino part is
proportional to $\hbar$, and thus can be thought of as a quantum
mechanical effect. It will turn out that the Wess--Zumino terms do
not live in eleven dimensions, 
in a sense that we will explain later, but
rather in twelve dimensions.

We still need to explain how the fields $\newB, \newH$ and $\tC$ are
related.

The worldvolume field strength $\newH \equiv  \myiota^*C - \dd \newB$ produces a
coupling of the bulk field $C$ and the five-brane worldvolume
2-form field $\newB$
\cite{Witten:1997hc,Pasti:1997gx,Aganagic:1997zq,Cederwall:1997gg}.
One way of seeing this is to write the Bianchi identity of the
worldvolume string flux {\it e.g.}~as in
Ref.~\cite{Witten:1996em}. The pull-back $\myiota^*$ in this definition
is that of the inclusion $\myiota : W_6 \hookrightarrow X $. Concretely,
if $X^\mu=X^\mu(\sigma)$ are the local coordinates on $X$ and
$\sigma^i$ those on $W_6$ we have
\be
\newH(\sigma)_{ijk} &=& - 3 \partial_{[i} \newB(\sigma)_{jk]} +
\frac{\partial X^\mu}{\partial\sigma^i} \frac{\partial
X^\nu}{\partial\sigma^j} \frac{\partial
X^\kappa}{\partial\sigma^k} ~ C(X(\sigma))_{\mu\nu\kappa}\ ,
\ee
where antisymmetrisation is performed ``strength one,'' {\it
i.e.~}$A_{[i}B_jC_{k]}=\ft16(A_iB_jC_k\pm 5 {\rm\ more\ terms})$.
In what follows, the distinction between fields that depend on the
spacetime coordinates $X^\mu$ and those that depend only on the
five-brane worldvolume coordinates $\sigma^i$ such as $C(X)$ resp.~$\newB(\sigma)$
is crucial.

An efficient way to keep track of this distinction and its physical repercussions is to
formulate the theory in the relative cohomology of the five-brane.
In relative cohomology, the pair $(C,\newB)$ is the potential of the
``relative'' field strength\footnote{We denote definitions
by``$\equiv$". Bianchi identities and  equations of motions are
denoted by ``$=$".}
\be
\dd (C,\newB) &\equiv& (G, \newH) = (\dd C, \myiota^*C - \dd \newB) \label{defGH}~.
\ee
The Bianchi identities then reduce to $\dd (G,\newH) =0$. They can be
violated by sources
\be
\dd(G,\newH) &=& \Big(\kappa \tilde T_6 ~ \delta(W_6), ~ T_{2\hookrightarrow
6} ~ \delta(U_2) \Big) \label{biGH}~,
\ee
where $U_2$ is some string worldsheet embedded within the worldvolume 
of the five-brane worldvolume $W_6$, and $\tilde T_6$ and
$T_{2\hookrightarrow 6}$ are normalisation coefficients. Gauss' law plus the 
BPS condition for the static five-brane requires
\be
\tilde T_6=T_6\ ;
\ee
we will fix the value of $T_{2\hookrightarrow 6}$ later.

In order to be able to treat magnetically charged configurations, we
should introduce dual field strengths into the formalism.
For the five-brane, the definition of an action is rather subtle
\cite{Witten:1997hc} because of the fact that the 3-form field strength
needs to be self-dual. However, as we are principally dealing with
Chern--Simons terms we shall be able to side-step these problems, and
self-duality only means for us that we do not need to introduce
another field strength on the worldvolume. The self-duality
equation
\be
\newH & = & -\frac{\newl}{\newe} *\newH \label{sdH}
\ee
will be imposed only after all variations of the action have been calculated.

The magnetic dual of the 3-form field $C$ is a 6-form $\tC$, and
we shall associate to it the 7-form field strength
\be
\tG &\equiv& \dd \tC + \half C \wedge G \label{deftG}
\ee
The dual gauge fields $\tC$ and $C$ are related by the duality
equation
\be
 \tG & = \frac{1}{\newg} *G & \label{sdG}~.
\ee
In fact, given $C$, it is possible to view this equation as the
definition of $\tC$, up to a gauge transformation.

We are now in a position to read off the gauge transformation
rules. From (\ref{defGH}) it is evident that the gauge fields $\newB$
and $C$ can be shifted by a closed term in relative cohomology;
The definition of $\tG$, however, warrants 
a more general shift; in the notation of App.~\ref{Relative}, we have
\be
\delta (C,\newB) &=& (\Lambda_3, \lambda_2 ) \label{galg1} \\
\delta \tC  &=& \Lambda_6 + \half \Lambda_3 \wedge C\ ,
\ee
where $(\Lambda_3, \lambda_2) \in H^3(X,W_6;\RR)$ and $\Lambda_6
\in H^6(X,\RR)$, so that the closure conditions boil down to
\be
\dd (\Lambda_3, \lambda_2 ) &=& (\dd \Lambda_3, \myiota^* \Lambda_3 -
\dd \lambda_2 ) = 0 \\
\dd \Lambda_6 &=& 0 ~.  \label{galg4}
\ee
These 3-form gauge transformations can be {\em large} outside the
five-brane; they reduce to small gauge transformation on the
five-brane worldvolume. In this sense they respect the five-brane
structure. Notice that the gauge transformations $\delta \newB =
\Lambda_2$ for closed 2-forms $\Lambda_2$ are just  special cases
of the above, {\it i.e.}\  $(\Lambda_3, \lambda_2 ) = (0, \Lambda_2)$.

The equations of motion become, {after} the  use of the
duality equations (\ref{sdH}) and (\ref{sdG}),
\be
 \frac{1}{\kappa} ~ \dd (\dd C) & = & T_6 \delta(W_6) \label{Ceom} \\
 \frac{1}{\kappa} \Big[ \dd (\dd \tC) - \half C \wedge \dd G \Big] &
= & T_3 \delta(W_3) + 2\newe \newH \wedge \delta(W_6)  \nonumber  \\
& &  - G \wedge \delta(2\newf W_7 + \newq \partial W_8)\label{tCeom}  \\
& &  - \kappa ~ \newq T_6 \delta(W_6) \wedge \delta(W_8) \nonumber~.
\ee
For the meaning of expressions of the form ``$2\newf W_7 + \newq
\partial W_8$" that mix real numbers and surfaces, see the end of App.~\ref{Currents}. 
The exterior derivative is nilpotent when acting on well-defined differential forms.
However, the gauge fields $C$ and $\tC$ fail to be well-defined exactly at brane
worldvolumes, so that ``$\dd^2$'' can produce the expected delta-function singularities.
In thinking of this it is useful to keep the Dirac monopole in mind: there, the field
strength is proportional to the volume form
$\Vol(S^2) = * \dd r^{-1}$ of the two-sphere surrounding the monopole, which is harmonic
outside the monopole. We shall continue to calculate\footnote{Other approaches to dealing
with this phenomenon are {\em either} explicitly to redefine $C=-\kappa T_6 \delta(W_7)+
\hat{C}$, where $ \delta(W_7)$ is defined in (\ref{deltaw7}), and calculate with the
well-defined form $\hat{C}$ as in Ref.~\cite{deAlwis:1997gq}, {\em or} to cut out a
tubular neighbourhood around the brane worldvolume and otherwise to modify every
appearance of the $C$-field to $\hat{C}-\sigma_3$ in the bulk as in
Ref.~\cite{Freed:1998tg}, {\em or} to calculate with Chern kernels as in
Ref.~\cite{Lechner:2001sj}.} with these gauge fields.

By requiring that these equations of motion be invariant under
gauge transformations, one finds relations between the
various coefficients ($\newa,\newe$ and $\newf$) and charges $T_3$ and $T_6$.
In doing this, it also turns out that the seven dimensional surface $W_7$, for
$\newf\neq 0$, becomes the Dirac surface of the magnetic five-brane source,
$\partial W_7 = W_6$.

The 2-form field equation and the Bianchi identity
\be
- \newl \dd *\newH - \newe~ \myiota^*G &  =  & \newk
~\delta_6(W_2) \label{Beom} \\
 - \dd \newH + \myiota^* G & =  &   T_{2\hookrightarrow 6}   ~ \delta_6(U_2)
\label{Bbia}
\ee
are actually not independent in a  supersymmetric theory: since the
3-form field strength is actually self-dual in these  supersymmetric models, 
$\newH = -\frac{\newl}{\newe} *\newH$, the two equations (\ref{Beom})
and (\ref{Bbia}) are in fact the same. Taking exterior derivatives
of (\ref{Bbia}) one finds ({\it cf.~}App.~\ref{Currents}) that
the worldvolume strings acting as sources of $\newH$ are closed.

The next step is to guarantee that the action is invariant at least under
small gauge transformations. This can be easily done and leads to
restrictions which, together with the above results, can be
summarised as
\be
T_3 &=& -\newa = 2\newk \label{con1} \\
T_6 &=& -3(\newe+\newf)  \qquad  \text{and}  \quad \newb = 0 \label{con2}  \\
T_{2\hookrightarrow 6} &=& - \frac{\newk}{\newe} \label{con3}~.
\ee
The surfaces $W_2=U_2= \partial W_3$ and $W_6 = \partial W_7$ are
without boundary. We have assumed from the start that there is no
Ho\v{r}ava--Witten boundary, {\it i.e.~}$\partial X =0$.

\subsection{Tensions of brane intersections}
\label{Thom}

Brane intersections are stable if they preserve some unbroken
supersymmetry \cite{Strominger:1995ac,Townsend:1997em,Chu:1998pb}.
For this to happen, a general requirement is the presence of certain
worldvolume fields that can support solitons at the
intersections. For instance, if a $p$-brane and a $q$-brane
intersect over a worldvolume $k$-brane, we need for stability to
have $k$-brane solitons in both the $p$- and the $q$-brane
worldvolume theories. By considering the charges of these
worldvolume solitons as Thom classes of the normal bundles, one
finds \cite{Papadopoulos:1999bz} the consistency condition
\be
T_{k \hookrightarrow q} T_q = T_{k \hookrightarrow p} T_p
\ee
in an obvious notation. This is exactly the situation we have been
talking about in the case $q=3,p=6,k=2$.

The only tension that we have not actually encountered yet is the tension of the
worldvolume string seen as a domain wall on the membrane, $T_{2
\hookrightarrow 3}$. Using Eqs (\ref{con1},\ref{con2},\ref{con3}) 
together with Eq.~(\ref{con4}), which will be found later in Sec.~\ref{Strings}, 
one finds that this tension does not appear as an independent parameter, and one has $T_{2
\hookrightarrow 3}=1$.

\section{Gauge invariant action}
\label{Strings}

The goal in this section is to turn the action $S$ into a machine that,
given a specific field-configuration, produces a well-defined complex
phase $\exp iS$ when it is evaluated on any collection of branes
$\{W_p\}$. We then say that $S$ is a differential character
\cite{CS} that takes its values in $\RR/2\pi\ZZ$.

What we need to do, more specifically, is to find a way to show
that any gauge transformation just amounts to shifting $S$ by
multiples of $2\pi$. One way of doing this is to write the action
in terms of quantities that are explicitly independent of the
choice of gauge, {\it i.e.~}field strengths. This is, of course, a
standard procedure for constructing Chern--Simons functionals,
{\it cf.~}App.~\ref{CSfnals}.

\subsection{The 12D action}
\label{Lift}

\begin{figure}[ht]
\begin{center}
\includegraphics{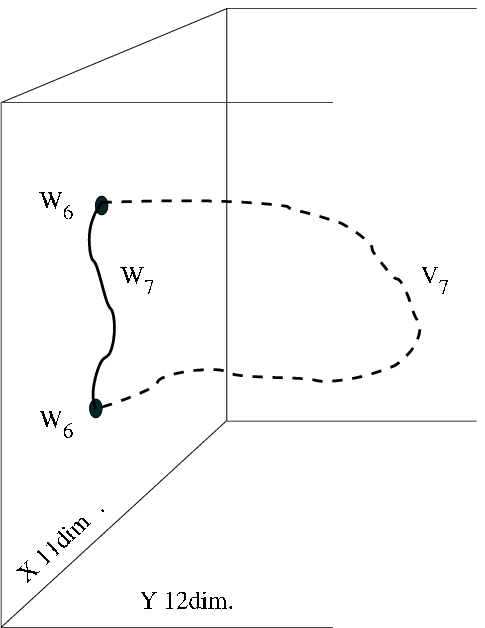}
\mycaption{$W_7$ and  $V_7$ have the five-brane worldvolume 
$W_6$ as a common boundary. 
The difference between them is that $W_7$ lies entirely in the 11D bulk $X$, 
whereas $V_7$ meets $X$ transversely from $Y$. The two components of $W_6$ serve
here to emphasise that the five-brane worldvolume has to be a boundary; of course in
practice $W_6$ can be connected.} 
\label{fig1}
\end{center}
\end{figure}

\begin{figure}[ht]
\begin{center}
\includegraphics{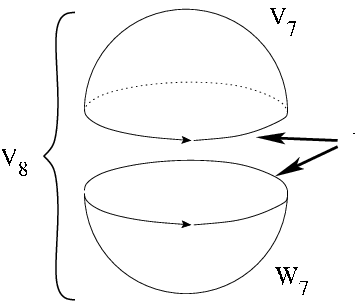}
\mycaption{Since $W_7$ and  $V_7$ have the same boundary, 
they can be glued together to form a closed surface, $\partial V_8=W_7-V_7$.} 
\label{fig2}
\end{center}
\end{figure}

Since the 11D spacetime $X$ is taken to be closed, it is a boundary of
some 12D spacetime $Y$. It can even be assumed that both are equipped
with compatible spin structures \cite{Witten:1985bt}. Let us
denote (with some abuse of notation) the inclusion of this boundary also by 
$\myiota : X \hookrightarrow Y$. In attempting to write the action in terms
of 12D quantities, we have to lift the
delta-function $\delta(W_6)$ onto $Y$. It is actually also true
that the cohomology classes $H^4(X, \ZZ)$ extend to classes in
$H^4(Y, \ZZ)$ \cite{Witten:1985bt}; here we just need to know that
there exists a surface $V_7 \subset Y$ such that the
delta-function on it pulls back to the five-brane delta function:
\be
\myiota^* \delta_Y(V_7) &=& \delta_X(W_6) \label{defV7}\ .
\ee
This means that $\partial V_7 \cap X = \partial W_7 = W_6$, as sketched in 
Figs \ref{fig1} and  \ref{fig2}. As $V_7$ has a boundary, 
the delta-function $\delta_Y(V_7)$ is not a closed differential form; 
since it fails to be so only by a piece that pulls 
back to zero in eleven dimensions -- {\it cf}.~App.~\ref{Currents} 
and Eq.~(\ref{fullStokes}) therein -- the 
exterior derivative kills consistently both sides of  
Eq.~(\ref{defV7}).
Gluing $V_7$ together with $W_7$ along their common boundary 
$W_6$, we get a closed
$d=7$ surface in $Y$, namely $W_7 \cup -V_7$ as sketched in Fig.~\ref{fig2}. 
The minus sign
denotes a reversal of orientation. Indeed,
if we now calculate the boundary, we get $\partial(W_7 \cup -V_7)
= W_6 -W_6 = 0$. It is convenient to think of unions as the addition
of sets. In fact, in the present context it makes sense to add and
subtract sets with arbitrary real coefficients, {\it
cf.~}App.~\ref{Currents}. We can therefore rewrite $W_7 \cup -V_7$ as
$W_7-V_7$, and we give the closed $d=7$ surface thus constructed the
name $\partial V_8$.

A similar treatment is necessary for $\delta_6(U_2)$, which is a
delta-function on the five-brane worldvolume $W_6$. Denoting the
inclusions again, abusively, by $ \myiota : W_6 \hookrightarrow W_7$, we
introduce a three-dimensional surface $U_3 \subset W_7$ such
that
\be
\myiota^* \delta_7(U_3) &=& \delta_6(U_2) \label{defU3}~.
\ee
Similarly to the above, its boundary is $U_2=W_2$ and we can glue
it together with $W_3$ and call the thus constructed closed
three-cycle $U_3-W_3 = \partial U_4$.

We shall assume that the four-surface $U_4$ and the
eight-surface $V_8$ actually exist. This means that
\be
\partial(V_8,W_7) &=& (V_7,W_6)  \label{rel76} \\
\partial(U_4,U_3) &=& (W_3,W_2) \label{rel32}
\ee
correspond to trivial classes in $H_7(Y, W_7)$, resp.~$H_3(Y,
W_7)$. However, this does not imply that the classes 
\be
[(V_7,W_6)] &\in&  H_7(Y,W_6) ~, \\ {} 
[(W_3,W_2)] &\in& H_3(X,W_6)
\ee 
are trivial, because cycles modulo boundaries in $W_6$ is a larger space 
than that of cycles modulo boundaries in $W_7$. 

Note also that this does not necessarily mean that spaces in which 
the above mentioned $H_7(Y, W_7)$ and $H_3(Y,W_7)$ homology groups
are nontrivial are inconsistent
backgrounds for wrapping branes.\footnote{Note that in
Ref.~\cite{Figueroa-O'Farrill:2001kz} this could not be avoided,
however, as nontrivial homology does obstruct the 
single-brane configuration studied in 
Ref.~\cite{Figueroa-O'Farrill:2001kz} 
on a compact space, specifically a Lie
group.} Rather than a restriction on cohomology of the background, this is
a restriction on the configuration.

\begin{figure}[ht]
\begin{center}
\includegraphics{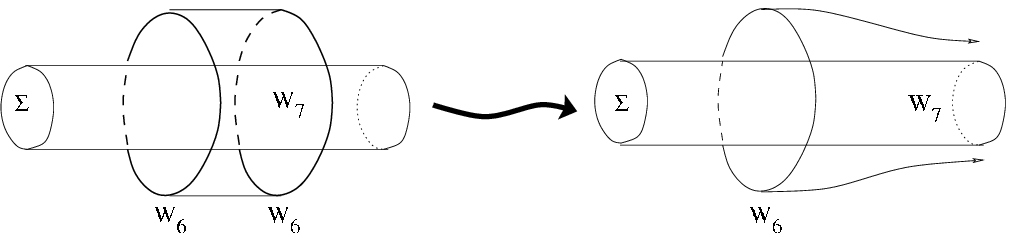}
\mycaption{Brane worldvolumes $W_6$ wrapped around non-contractible cycles 
$\Sigma$ need to come in pairs, so that a connecting homotopy $W_7$ 
exists. In a noncompact space one of these can be taken to be at infinity, 
which produces the standard Dirac surface picture.} 
\label{fig9}
\end{center}
\end{figure}

In this discussion, as we have already sketched in Fig.~\ref{fig1}, 
the notation $W_6$ does not necessarily refer
to an isolated brane, but rather can refer to a collection of them.
The assumption of topological triviality made here is merely a
statement of how these branes relate to each other. With this
assumption, one cannot consistently describe an isolated five-brane
in a compact space; there has to be at least one more brane
present, such that there exists a homotopy between the two. This
makes the homology class trivial, although the  branes  that we use to
represent it are still physically nontrivial. Moving a five-brane worldvolume 
along this homotopy would then sweep out the worldvolume of the
pertinent Dirac surface, $W_7$. In a noncompact space the extra
brane worldvolume can be taken to infinity; this is indeed Dirac's classical
construction, sketched in Fig.~\ref{fig9}. 
In other words, the brane charge has to have
somewhere to flow.

The only genuine restrictions on the model that we make are the assumptions
that $W_6$ and $X$ have no boundaries. Note that this assumption does not 
imply that $W_6$ or $X$  has to be compact; open spaces without boundary, 
{\it e.g.}~$\RR^{11}$, are also acceptable.
Even these restrictions could perhaps be relaxed by considering 
manifolds with edges, such as in extended topological quantum field theory.

We are now ready to write down the action\footnote{
Note that this assumes that the worldvolume field $\newH$ can be extended from
$W_6$ onto $W_7$; There may be obstructions to doing this.
Note also that we have set $1/\newg=-\newl/\newe=+1$. 
These factors can be easily reinstated in front of
the kinetic terms.} in Dirac surface form:
\be
 S_\gauge &=& 
       \int_X \frac{1}{2\kappa}~ G \wedge * G 
     + {T_3} G \wedge \gravano_7(R) 
     - \int_Y  \frac{1}{6\kappa} G \wedge G \wedge G  
       \nonumber \\
& &  - \half T_6 \int_{(V_8,W_7)} 
       \Big(  G \wedge G - 2\kappa T_3 ~\gravano_8, 
               ~ \newH \wedge \myiota^* G \Big)    
       \label{strings} \\
& &  + \frac{T_6}{4}  \int_{W_6} \newH \wedge *\newH 
     + {T_3}  \int_{U_4} G 
     - \frac{T_3}{2} \int_{U_3} \newH  ~.  
       \nonumber
\ee 
Here we have made use of the facts that $\partial(U_4,U_3) =
(W_3,W_2)$ and that
\be
-   \int_{(W_3,W_2)} (C,\newB) &=&   + \int_{(U_4,U_3)} (G,\newH)~.
\ee
As $U_3 \subset W_7$ we may think of $U_3$ as a surface 
traversing through $W_7$, in contrast to $W_3$, which does not need
to lie within $W_7$. One could perhaps think of $W_3$ as an
electric and $U_3$ as a magnetic Dirac surface from the point of
view of the self-dual theory on the five-brane.

We have also introduced in (\ref{strings}) the gravitational corrections
$\gravano_8,\gravano_7$ of Appendix \ref{Gravitational}. The purely
gravitational integral of $\gravano_8$, defined later in (\ref{omega8}),
is added here for
convenience, although the evidence\footnote{This is the same correction, in
effect, as in \cite{deAlwis:1997gq}. It yields
the Bianchi identity for $\tG$ so that independence of the action
from the choice of $V_8$ becomes manifest. It also  forces one to
fix the homology class of $V_8$ such that $\partial V_8 =
W_7-V_7$. See also the comparison to previous 
work immediately below.} in its favour will surface later after calculating 
the contributions from integrals over $V_8$ in 
(\ref{v8contrib}).  The gravitational 
corrections lead to modifications in the definition of $\tG$ as
well:
\be
\tG &\equiv& \dd \tC + \half C \wedge G - \kappa T_3 \gravano_7~.
\ee

The action (\ref{strings}) is essentially the same as the one proposed in
Eq.~(7.25) of Ref.~\cite{deAlwis:1997gq}. The Dirac surfaces appear
there through explicit delta-functions $\theta_4$ and
$\theta_7$. The only differences\footnote{The comparison makes use
of the assumption that these delta-functions restrict onto the world
volumes $W_7$ and $U_4$, respectively. Then the calculation
reduces to writing $G=G_{\text{de Alwis}}+ \kappa T_6 \theta_4$
and expanding. As \cite{deAlwis:1997gq} disregards products of
delta-functions we are at liberty to do so here as well.} with respect to
\cite{deAlwis:1997gq} are in the identification of the physical
membrane tension, and in the purely gravitational term
$\int_{V_8}\gravano_8$, as opposed to the term $\int_{V_7}\gravano_7$ given in 
Ref.~\cite{deAlwis:1997gq}. As
the membrane tension in the equations of motion derived from the
two actions is nevertheless the same, a factor of two difference
in the integral over the membrane worldvolume reduces purely to a self-duality
related convention. 

Two surprising features of the action (\ref{strings}) are that there is no
explicit $\tC$ dependence and that the coupling to $C$ and $\newB$ fields is
not through the relative pair $(C,\newB)$. The former fact is
a reflection of the {\em locality} of the action and the latter is 
a reflection of the nested structure of the geometry together with consequences of the
self-duality of $h$.

We still need to show that this form of the action is independent
of the choice of the Dirac surfaces. We will do so in Section
\ref{Dependence}.

Let us define the violated Bianchi identities in the bulk as
\be
\DD (G,  \newH) &\equiv& (\dd G, \myiota^*G - \dd \newH ) \nonumber \\
&=& \Big( \dd (\dd C), \dd (\dd \newB) \Big)  \\
\DD \tG &\equiv& \dd \tG - \half G \wedge G + \kappa T_3~ \gravano_8 \nonumber \\
&=& \dd (\dd \tC) - \half C \wedge \dd G~.
\ee
The bulk equations of motion/Bianchi identities then take the form
\be
 \frac{1}{\kappa} ~ \DD G & = & T_6 \delta(W_6)  \label{seomC} \\
 \frac{1}{\kappa} ~ \DD \tG &
= &   T_3 \delta(U_3-\partial U_4) +
T_6 \newH \wedge \delta(W_6)  \nonumber  \\
& &  + T_6 G \wedge \delta(V_7 + \partial V_8 - W_7)
\label{seomtC} ~.
\ee
The last term in (\ref{seomtC}) actually drops out upon use of the surface relation $V_7 +
\partial V_8 - W_7 =0$, which we shall obtain shortly.

From the above discussion, we see that the proper global formulation 
of the theory as given in (\ref{strings}) 
guarantees the absence of awkward terms
such as the last two terms in (\ref{tCeom}). The
magnetic sources are then five-brane worldvolumes $W_6$, as expected, and the
electric sources are the membrane worldvolumes $W_3$. Furthermore, membrane number
can turn into worldvolume string flux  through the coupling $\newH
\wedge \delta(W_6)$ with the correct multiplicity.

In writing down the action (\ref{strings}) we have actually used
some facts that strictly speaking follow only from inspecting the equations of motion
(\ref{seomC},\ref{seomtC}). In particular, we have used the fact that 
\be
T_6=-2\newe ~. \label{con4}
\ee
This result, which was anticipated in the considerations of Sec.~\ref{Thom}, follows from
taking the exterior derivative of (\ref{seomtC}) and comparing the
terms proportional to $\delta(W_6)$ with (\ref{Beom}). The
cancellation of the rest of the terms when taking the exterior
derivative of (\ref{seomtC}) requires\footnote{There is a
possible pitfall here that one needs to avoid. Another, although erroneous, way of thinking
would be to consider that the five-brane tension
$T_6=-3(\newe+\newf)$ as derived from the original action
(\ref{bulk}--\ref{wz}) might differ from the tension
$T_6'=-2\newe$ arising from (\ref{strings}). This would be  tantamount
to claiming that two actions for one and the same theory
could somehow be associated to different phases. This would also lead to
additional difficulties in finding the correct Dirac quantisation
conditions \cite{proc}. In particular, this
would require one to postulate fractional charges and substructure,
thus conjuring up a picture of hadronic and quark-like branes, as
raised as a possibility in Ref.~\cite{proc}.} that the combination of
surfaces $V_7 + \partial V_8 - W_7$ appearing in the
first delta-function in (\ref{seomtC}) vanish identically. This then fixes the choice of
the eight-dimensional Wess--Zumino term to be $\newq W_8 = -(T_6/3)
V_8$. Conversely, the presence of $W_7$ and $W_8$ in Eq.~(\ref{wz}) {\em requires} 
that there exist some surface $V_8$ so that the boundary relation (\ref{rel76}) holds. 

\subsection{The Hopf--Wess--Zumino term}
\label{hwz}

Notice that in formulating the theory in this way,
12D data has to be included in the model from the
very beginning. In the topologically trivial case, 
the original Wess--Zumino term therefore
becomes 
\be
S_{\wz}&=&
 \newf \int_{W_7} C \wedge G  + \frac{\newq}{2} \int_{W_8} G \wedge G \nonumber \\
&=& \frac{T_6}{6} \int_{W_7 - \partial V_8} C \wedge G \nonumber \\
&=& \frac{T_6}{6} \int_{V_7} C \wedge G \label{addV7} ~.
\ee 
Though the equalities here do not strictly hold in 
the topologically nontrivial case, 
the difference is of no consequence in the analysis 
performed above. 
Therefore, we can {\em either} start from the action 
that involves $W_7$ and $W_8$, and discover 
the topological constraint  (\ref{rel76}) when 
investigating the invariance properties of 
the full interacting theory {\em or} from the Wess--Zumino 
term (\ref{addV7}) and the definition 
(\ref{defV7}). This means that the Wess--Zumino term should 
be really seen as inherently 12D data. 

This term is also quite closely related to the 
Hopf--Wess--Zumino term that was proposed in 
\cite{Intriligator:2000eq} to
ensure correct anomaly cancellation in holography in 
the case of a stack of five-branes. As discussed in more detail in App.~\ref{CSfnals}, 
the field strength behaves near the five-brane worldvolume as $G = \kappa T_6 \Omega_{\sst 4} +\hat{G}$, 
so that when the smooth part vanishes, $\hat{G}=0$, the Wess--Zumino term becomes, using 
Eqs (\ref{quant1},\ref{quant2}) to be found in the next section 
\be
S_{\wz}&=& \frac{\kappa T_6^3}{6} \int_{V_7} \Omega_3 \wedge \Omega_4 = 
           \frac{2\pi}{6} \int_{V_7} \Omega_3 \wedge \Omega_4 ~.
\ee
For $V_7=S^7$ this is precisely the Hopf-invariant of a class in 
$\pi_7(S^4)$ suggested in \cite{Intriligator:2000eq} for 
the anomaly deficit $c(G)-c(H)=1$. Though this involves extrapolating 
the gauge group to a single brane configuration, the above should 
serve as a microscopic supergravity derivation of the interaction 
term conjectured in \cite{Intriligator:2000eq} on grounds of 
duality principles.

\section{Consistency conditions, gauge and charge lattices}
\label{Consistency}

In this section we will check that the Dirac surface form of the action
is well-defined modulo $2\pi$, and will derive quantisation conditions
for fluxes, tensions, and large gauge transformations.

\subsection{Dirac quantisation}
\label{Dirac}

We first show that the fluxes $G$, $(G,\newH)$ and $\tG$ are
quantised in a certain sense, even if not all of them are closed classes in
cohomology. This will be sufficient to show that the large gauge transformations
take  their values in a certain integer lattice.

\begin{figure}[ht]
\begin{center}
\includegraphics{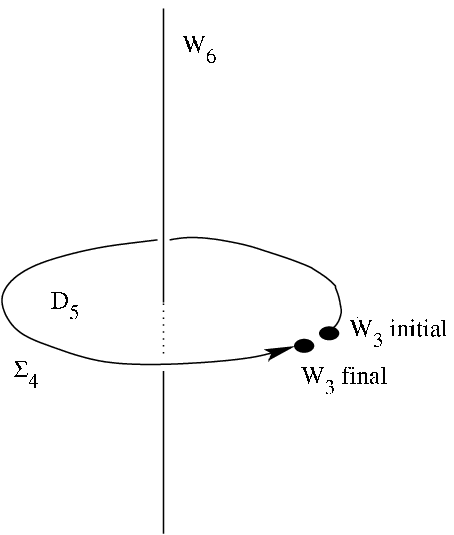}
\mycaption{The membrane worldvolume $W_3$ can be transported around 
a five-brane worldvolume $W_6$ along a closed path $\Sigma_4$. 
This path can be filled in to form a disc $D_5$ which the 
five-brane worldvolume $W_6$ intersects once transversely.} 
\label{fig3}
\end{center}
\end{figure}

The first check on the partition function is to see what happens
when one transports a membrane worldvolume adiabatically around a five-brane
source worldvolume as sketched in Fig.~\ref{fig3}. 
Let us suppose first that the membrane and the five-brane
are well separated and that the former does not end on the
latter. This leads to the standard Wu--Yang derivation \cite{Wu:qk} of the
Dirac quantisation condition \cite{Dirac}. The phase of the membrane coupling,
when transported along a curve $\Sigma_4$, is
\be
T_3 \int_{W_3^\mathrm{fin}} C - T_3 \int_{W_3^\mathrm{ini}} C =
T_3 \int_{\Sigma_4} G \stackrel{!}{=} 2\pi ~.
\ee
This tells us that the integrals of $\frac{T_3}{2\pi} G$ are integers\footnote{It is
simplest to  think here of compact closed surfaces, which  arise naturally if we apply
the Wu--Yang argument  to branes that are localised in time. We could equally well,
however,  have an infinite time direction on the membrane  worldvolume, so that the
surfaces then would be closed but noncompact.}  when calculated over surfaces that
neither have a boundary nor intersect the five-brane source worldvolume itself,
\be
\left[ \frac{T_3}{2\pi} G\right] \in H^4(Y,\ZZ)\ . \label{flux1}
\ee
If we view $\Sigma_4 =
\partial D_5$ as a boundary of a five-disc and also use the equation of
motion (\ref{Ceom}), we obtain the standard Dirac quantisation
condition
\be
\kappa ~ T_3 T_6 = 2\pi \label{quant1} ~.
\ee

\begin{figure}[ht]
\begin{center}
\includegraphics{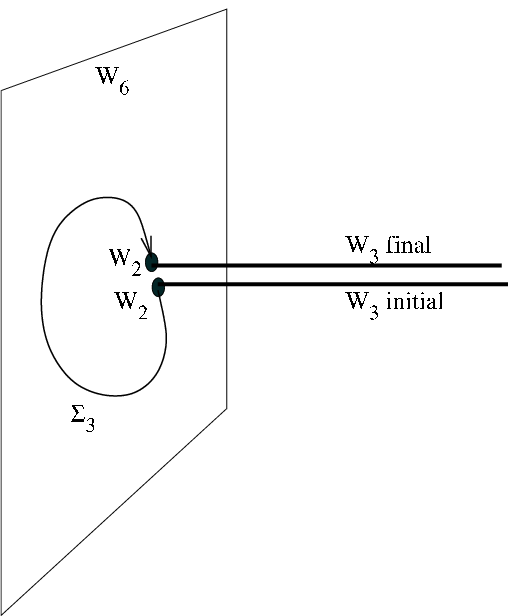}
\mycaption{Since the worldvolume strings $W_2$ 
are the membranes' endpoints, moving one 
means moving the other.} 
\label{fig4}
\end{center}
\end{figure}

\begin{figure}[ht]
\begin{center}
\includegraphics{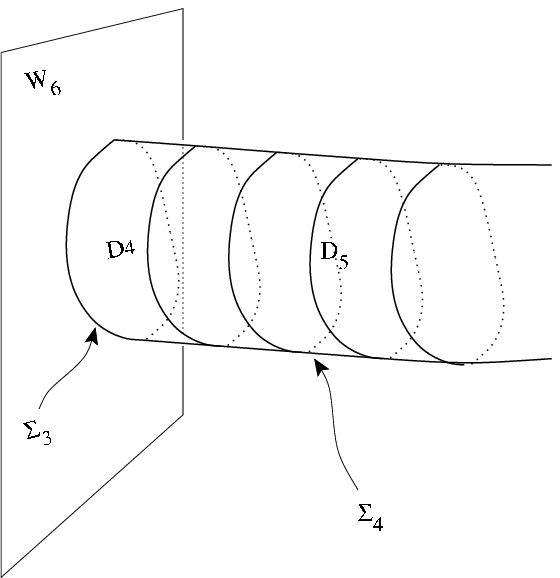}
\mycaption{$\Sigma_3$ is the path that the string worldvolume $W_2\subset W_6$
adiabatically traverses, and $\Sigma_4$ is that traversed by the membrane worldvolume.
$\Sigma_4$ is therefore the surface of a (hollow) cylinder that ends on $\Sigma_3$ in
$W_6$.  If we fill in the latter to make a 4-disc $D_4$ and add this to $\Sigma_4$ we 
get a capped cylinder, which can then be filled in to yield a 
solid cylinder $D_5$.} 
\label{fig5}
\end{center}  
\end{figure}

It is also possible to use this same argument even in cases where the membrane ends on
the five-brane; this yields a quantisation condition for the tension $T_{2\hookrightarrow
6}$ of the self-dual strings in the five-brane worldvolume. The argument is more subtle,
however, because of the self-duality. It is useful to think of any string source $W_2
\subset W_6$ as a sum of an electric and a magnetic source, which are interchanged by
self-duality.\footnote{As usual with issues relating to the quantisation of self-dual
fluxes ({\it cf.}~Ref.~\cite{Henningson:1999dm,Witten:1997hc}), the matter calls for
caution. In the calculation of a five-brane partition function we actually sum over only
a specific choice of polarisations of the dyonic flux lattice, say over the electric
fluxes. What we are saying here is that summing over all of the fluxes and dividing by
two in the current couplings gives the correct result presented here.}  They are at the
boundary of a bulk membrane worldvolume $W_3$, which we consequently have to move in an
adiabatic fashion along with the attached worldvolume strings as well, {\it cf.}\ 
Fig.~\ref{fig4}.  The part of the action (\ref{cs}) that is affected by the adiabatic
argument is then
\be 
-T_3 \int_{(W_3,W_2)} (C,\half \newB) \label{naiverelcur}
\ee
Since membranes are not self-dual, the adiabatic argument yields
the change of phase 
\be
T_3 \int_{(W_3,W_2)_\mathrm{fin}} (C, \newB) - T_3   
\int_{(W_3,W_2)_\mathrm{ini}} (C, \newB) = T_3 \int_{(\Sigma_4,
\Sigma_3) } (G,\newH)  \stackrel{!}{=} 2\pi~. 
\ee 
Note that the factor of $\ft12$ has now vanished as anticipated above.  
Here $(\Sigma_4,\Sigma_3)$ describes the path along which we
transport the combined membrane -- string system, {\it cf.}\   Fig.~\ref{fig5}.  

On the level of cohomology, the above considerations mean that the class $\frac{T_3}{2\pi}
(G,\newH)$ is integral,
\be
\left[\frac{T_3}{2\pi} (G,\newH)\right] \in  H^4(Y,W_6;\ZZ)\ .\label{flux2}
\ee
We can further fill in the cylinder shown in Fig.~\ref{fig5} so that
$(\Sigma_4,\Sigma_3) = \partial(D_5,D_4)$, and we can use the equation of
motion (\ref{Beom}). Now it is important to notice that 
the integral 
\be
\int_{D_5} \delta_X(W_6) = 0
\ee
vanishes because the five-volume $D_5$ meets the five-brane worldvolume $W_6$ only along a dimension-four surface  
$D_4$. Indeed, the current $\delta_X(W_6)$ involves differentials along the five
transverse  directions of the five-brane worldvolume, whereas the integration domain $D_5$
only includes  one of these directions.  We get therefore a quantisation condition that
involves only terms relating to the worldvolume string and the membrane,
\be
T_{2 \hookrightarrow 6} T_3 = 2 \pi~.\label{stringmembrel}
\ee
Now recall that $T_{2 \hookrightarrow 6}=T_3/T_6$ was related to the bulk
tensions so that finally we have
\be
(T_3)^2 &=& 2\pi T_6 \label{quant2}~. 
\ee 

We will rederive this in different ways later on. The above argument also serves to elucidate
the relationship between relative homology and Dirac surfaces. There are two observations to
be made at this point: 
\begin{itemize}
\item[1)] 
The results (\ref{quant1}) and (\ref{quant2}) follow also 
by requiring that the action (\ref{strings}) remain well-defined modulo $2\pi$ when
we distort $U_4$ or $U_3$ by a surface without a boundary. 
Here, in order to understanding the factor of $\ft12$ in front of the last term in
(\ref{strings}), one must remember that that the integral involves only half of the na\"{\i}vely
apparent fluxes by virtue of the self-duality of $h$.
\item[2)]  
The result (\ref{stringmembrel}) also follows from the quantisation of dyonic charges as
discussed in \cite{Bremer:1997qb}. There, it was noted that, in $D=4k+2$
dimensions and for Lorentzian signature, the generalised
Dirac--Schwinger--Zwanziger quantisation condition \cite{Schwinger,Zwanziger} 
involves a relative plus-sign, {\it i.e.}\ 
$e_1 g_2 + e_2 g_1 = 2\pi \kappa$. In the case of a self-dual flux we thus
have an extra factor of two, and so $e_1 e_2 = \pi \kappa$. 
\end{itemize}

Let us finally turn to the membrane flux $\tG$. Collecting
all the $V_8$ dependent terms in the action we get
\be
\int_{V_8} \cdots &=& -T_6 \int_{ V_8} \dd\tG+ T_6 \int_{V_8} \DD \tG \\
&=& -T_6 \int_{\partial V_8}  \tG + \kappa T_6  T_3  \int_{V_8}
\delta_Y(U_4) + \kappa (T_6)^2  \int_{V_8} \newH \wedge \delta_Y(V_7)
~. \ \ \label{v8contrib}
\ee
The last term is at least na\"{\i}vely trivial, because integrals over $\delta_Y(V_7)$ could
possibly contribute only when taken over domains that include the five normal directions of
$V_7$, whereas $V_8$ only includes one of them. This term may play a r\^{o}le if we
consider more complicated self-intersections of $V_7$ and $V_8$, but even then,
assuming that $\dd \newH=0$ and $\partial (V_8 \cap V_7) = 0$ and
using the above-found relations for the tensions, we see that this
term as well as the penultimate term contribute only integer
multiples of $2\pi$. The $V_8$ dependence of the ``$\mod
2\pi$''-action comes therefore entirely from the membrane flux
through the surface $\partial V_8 = W_7 - V_7$.

It does not make sense to require that the membrane flux $\tG$ be a class in integral 
cohomology, since it is not even closed when $G\wedge G$ or $\gravano_8$ are 
nontrivial. However, we have already seen above that even fluxes that are not generically closed, 
such as $\newH$, may nevertheless be associated to a quantisation condition in the full
interacting theory,  such as Eq.~(\ref{flux2}).  Furthermore, we must be able to show
that the action (\ref{strings}) is independent of the choice of  the Dirac surface $V_8$. This is
actually possible, and the proof proceeds in two steps  which we relegate to subsequent
sections. It turns out that  the $\dd \tC$ and the $\half G \wedge G - \kappa T_3 \gravano_7$
parts can be  discussed independently: the treatment of the former gives rise to quantisation of 
the six-form gauge transformations, as we will see at the end of Sec.~\ref{Lattice};
the latter will  require a full analysis of deformations of the Dirac surfaces, which 
we will make in Sec.~\ref{Dependence}. 

\subsection{The lattice of large gauge transformations}
\label{Lattice} 

The fact that the integral of $T_3 G$ over any four-sphere is an
integer multiple of $2\pi$ means that the allowed large gauge
transformations relating the values of the gauge field
on different hemispheres must also lie in integral cohomology,
\be
\left[ \frac{T_3}{2\pi}\Lambda_3 \right] &\in& H^3(Y,\ZZ)~.\label{gtquant1}
\ee
In more detail, we may divide an integral of $G$ over $S_4$ into integrals
over 4-discs $D,D'$ where the field strength can be trivialised
to $dC,dC'$. The gauge fields on $D$ and $D'$, with $\partial D=\partial D'=S^3$, differ by a
gauge transformation,
$C'=C+\Lambda$, so that
\be
\int_{S^4} G = \int_{D'} \dd C' + \int_{D} \dd C = \int_{S^3} C' -
C = \int_{S^3} \Lambda~.
\ee
Similarly, the flux quantisation condition (\ref{flux2}) implies 
\be
\left[  \frac{T_3}{2\pi}(\Lambda_3,\lambda_2)  \right] &\in&
H^3(Y,W_6;\ZZ)~. \label{gtquant2}
\ee
In Eqs (\ref{gtquant1}) and (\ref{gtquant2}) we have used the fact that 
the 4-form classes $[T_3 G/2\pi]$ extend as integral classes from $X$ to $Y$, as was 
mentioned in the beginning of Sec.~\ref{Consistency}. The large gauge transformations  
(\ref{gtquant2}) take values in the relative cohomology of the pair $(Y,W_6)$  rather than
in that of the pair $(Y,W_7)$ because the flux quantisation condition (\ref{flux2}) 
did not involve the Dirac surface $W_7$ in any way. 

A perhaps more direct way to see this is to notice that the 
worldvolume couplings 
\be
\int_{(W_3,W_2)} (C,\newB) \label{truerelcur}
\ee
remain invariant modulo $2\pi$ only if the above quantisation conditions apply.
The action (\ref{strings}) actually involves a slightly different combination 
of these terms, as pointed out in Eq.~(\ref{naiverelcur}), but as the discussion
there implied,  the coupling that takes self-duality properly into account is indeed of the form
(\ref{truerelcur}).

\begin{figure}[ht]
\begin{center}
\includegraphics{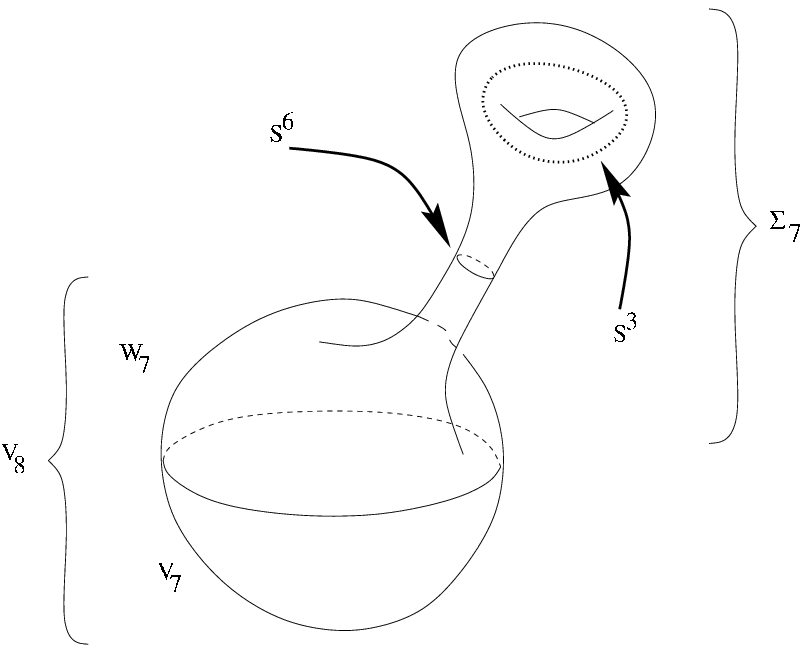}
\mycaption{Deforming $\partial V_8$, and therefore $W_7$ (or $V_7$) by adding a handle
$\Sigma_7$, after cutting out an original disk $D_7$.  The gauge field $\tC$ undergoes a
large gauge transformation somewhere in the neck, for  instance around the sphere $S^6$. The
handle may also contain noncontractible three-spheres that have an influence on the
M-theory gauge algebra.} 
\label{fig10}
\end{center}
\end{figure}

Even if there is no direct coupling to $\tC$ 
we will still need to restrict 
the large six-form gauge transformations to the lattice
\be
\left[  \frac{T_6}{2\pi}\Lambda_6  \right] &\in& H^6(Y,\ZZ) ~,  \label{gtquant3}
\ee
as was promised in the discussion below Eq.~(\ref{v8contrib}). 
To see this, consider deforming the surface $\partial V_8$ by inserting a handle 
without boundaries, such as sketched in Fig.~\ref{fig10}. It may well then happen that the
gauge field 
$\tC$ has to undergo a large gauge transformation $\Lambda_6$ 
somewhere around the neck of the deformation, say at the sphere $S^6$. 
More concretely, let us cut out from the original surface $\partial V_8$ some small 
7-disc $D_7$ with boundary $S^6$. We may then deform the cut surface by pulling the $S^6$
boundary some small distance away form its original location on the surface $\partial
V_8$ as sketched in Fig.~\ref{fig10}. We may next glue along  the boundary $S^6$ a handle
$\Sigma_7$. If  it happens that the 6-form gauge field on the original surface $\tC_0$  and on
the handle  $\tC_1$ differ by a large gauge transformation  
$\tC_0-\tC_1 = \Lambda_6$ then the  term 
\be
-T_6 \int_{\partial V_8}  \dd \tC 
\ee
in the $V_8$-part of the action (\ref{v8contrib}) is subject to a shift by the amount 
\be
-T_6 \int_{\partial V_8 - D_7} \dd \tC_0  -T_6 \int_{\Sigma_7} \dd \tC_1 = T_6 \int_{\partial D_7} \tC_0 - 
T_6 \int_{\partial \Sigma_7}  \tC_1 
= T_6 \int_{S^6}  \Lambda_6 \ .
\ee
Requiring the theory to be insensitive to this deformation gives rise to the integrality condition
(\ref{gtquant3}) above. 

Notice that it is important that the cut surface boundary modelled here by $S^6$ 
may be anywhere in space. In fact, had we restricted ourselves to comparing the values of
the 6-form gauge fields only on the original Dirac surfaces $W_7$ and $V_7$, the comparison
would have to be made at the joint boundary of $W_7$ and $V_7$, {\it i.e.}\ on the original
five-brane worldvolume $W_6$. This would then complicate the discussion that we shall give
of the gauge algebra in Sec.\ \ref{Algebra}. Indeed, were the gauge transformations in this
discussion always localised on $W_6$, then by $\myiota^*\Lambda_3=\dd \lambda_2$,
the wedge product of two 3-form transformations would give rise to a 6-form
transformation $\Lambda_6=\Lambda_3'\wedge\Lambda_3$ that is a trivial class when pulled
back to $W_6$, and consequently we would not obtain any constraint on the 6-form gauge
transformation lattice.

In conclusion, even though the 7-form flux does not itself need to be closed or integral, 
the transition functions of the 6-form gauge field, and consequently 
the 6-form gauge transformations, need to be integral (\ref{gtquant3}) in order to  
guarantee that the action is well-defined. 

\subsection{Dependence on Dirac surfaces}
\label{Dependence}

Guaranteeing independence of the theory from the choice of the surfaces
$Y$ and $V_8$ under shifts by closed surfaces boils down to requiring
that
\be
\left[\frac{1}{3! ~ \kappa} G^3\right]  \quad \text{  and   } \quad
\left[\frac{T_6}{2!} G^2\right]
\ee
are integral. The first of these two integrality conditions was found, in essentially
the same way, in Ref.~\cite{Duff:1995wd}; see also the appendix in
Ref.~\cite{deAlwis:1996ez}. Using (\ref{flux1}) this means
\be
\frac{(2\pi)^2}{\kappa} = (T_3)^3  \quad  \text{  and   }  \quad
2\pi T_6 = (T_3)^2 ~, \label{d1}
\ee
which is  equivalent to requiring that (\ref{quant1}) and
(\ref{quant2}) hold. The treatment of these formulas in the face of  
gravitational corrections would require a discussion of anomalies.

These quantisation rules are necessary, but not quite sufficient
to guarantee the independence of the theory from the choices of
Dirac surfaces that we have made, because they only guarantee 
the independence from the choice of the surfaces $Y$ and $V_8$. 
Showing independence also from the choices of $V_7$ and $W_7$, 
is further complicated because the Bianchi identity for $\newH$ 
given in (\ref{Bbia}) is obviously not of the kind that we see in 
standard de Rham cohomology. This issue can be settled by 
checking what happens to the action when we shift
all of the Dirac surfaces arbitrarily outside the sources 
(\ref{Beom},\ref{Bbia},\ref{seomC},\ref{seomtC}). 
This is essentially an exercise in 12D Euclidean geometry, but 
we give here an outline in algebraic form.

\begin{figure}[ht]
\begin{center}
\includegraphics{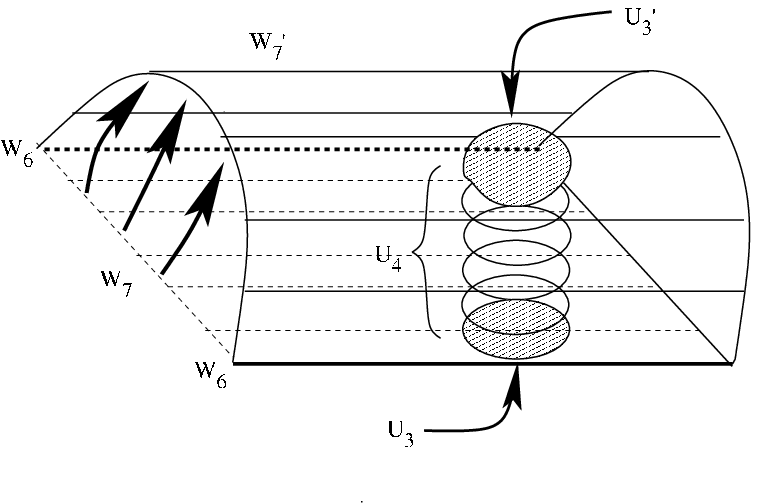}
\mycaption{Since $U_3$ was defined as a subset of $W_7$ whose boundary 
is the string worldsheet $W_2$, varying $W_7$ means varying $U_3$ while keeping 
the boundary fixed. Under 
smooth variations $U_3$ sweeps out a volume $U_4$.} 
\label{fig6}
\end{center}
\end{figure}

Quite generally, if we deform $W$ smoothly to $W'$, we sweep out thereby a
surface whose boundary is exactly $W'-W$ ({\it cf.}~Fig.~\ref{fig6}). 
As some surfaces lie by
definition within the worldvolumes of others, such as $U_3 \subset
W_7$, moving the latter moves automatically the former
as follows:
\be
Y' = Y + \partial Z & \Longrightarrow & V_7' = V_7 + \partial Z_8 \label{bound1}
\\
W_7' = W_7 + \partial D_8 & \Longrightarrow & U_3' = U_3 +
\partial D_4\ .
\ee

Apart from these standard Dirac surfaces, in the context of
relative cohomology we have also to deal with ``Dirac surfaces for
Dirac surfaces" such as $U_4$ and $V_8$. These are volumes whose
boundaries consist partially of physical branes and partially of
Dirac surfaces. As we do not know {\em a priori} what is physical
and what is not, we will have to vary these volumes arbitrarily
\be
V_8' = V_8 + T_8 & & \\
U_4' = U_4 + T_4 & &\ .\label{bound4}
\ee

The total shift in the action under (\ref{bound1}--\ref{bound4}) is
\be
 \frac{T_6}{2} \int_{D_8 - T_8 - Z_8 } G \wedge G
+ 2\pi \int_{T_8} \gravano_8 +  {T_3}   \int_{T_4 - D_4} G +
\frac{T_3}{2} \int_{(D_4, \partial D_4)} (G,\newH) ~. \nonumber \\
\label{delta}
\ee
The last term in (\ref{delta}) is trivial, because it is a pairing of a
relative cohomology class  with a trivial homology class; this is
why the factor of $\ft12$ here does not cause any further change in the result.
Na\"{\i}vely, it would seem to be necessary to require the other integration domains to be
empty. However, since it is sufficient to require these terms to integrate to multiples of
$2\pi$, we need these domains only to be without boundary, by virtue of the integrality
rule (\ref{flux1}):
\be
\partial (-D_8 + T_8 + Z_8 ) &=& 0 ~ , \label{bound5}\\
\partial T_8 &=& 0 ~ , \\
\partial T_4 - \partial D_4 &=& 0 ~ .\label{bound6}
\ee

Consequently, the integrality conditions (\ref{flux1}) guarantee
that the ambiguity in
the choice of Dirac surfaces indeed boils down to shifts in the
action by integer multiples of $2\pi$, as required. Note that for
this we need, in addition to the integrality of $G$ in
(\ref{flux1}),  to require also that
\be
[\gravano_8] \in H^8(Y,\ZZ) ~.
\ee
This condition is already familiar, however, from the membrane tadpole
cancellation arguments of \cite{Sethi:1996es,Witten:1997md}.

We can look at the above relationships between shifts in essentially
two ways. The first is by checking that the relationships (\ref{rel76},\ref{rel32}) found
between Dirac surfaces guarantee that the constraints (\ref{bound5}--\ref{bound6}) are
satisfied. This would mean that the Dirac surface form of the action (\ref{strings}) is a good
representation of the original form (\ref{bulk}--\ref{wz}) of the action. This is indeed the
case:
\begin{quote} 
{\em Proof:}
The fact that $T_8$ cannot have a boundary is
guaranteed by the fact that $W_7$ and $V_7$ fix the cobordism
class of $V_8$ once and for all. The requirement $\partial (-D_8 +
T_8 + Z_8 ) =0$ follows from requiring that $-W_7 + V_7 - \partial
V_8$ be preserved under deformations. And $\partial
T_4=\partial D_4$ follows from the construction $\partial U_4 =
U_3 -W_3$, which actually guarantees even that $T_4=D_4$.
\end{quote} 
The second way of looking at this is to note that one can show that the relations
(\ref{rel76}) and (\ref{rel32}) arise from (\ref{delta}), so that (\ref{strings})
carries enough information to tell us which combinations of the
many surfaces ($U_3,U_4,W_7,V_8,X,Y$) are actually physical, and
which ones are mere artifacts of the description, {\it i.e. }\ Dirac
surfaces: 
\begin{quote} {\em Proof:} The two last terms of (\ref{delta}) betray that the shift
\be 
\delta(U_4,U_3) &=& (D_4,\partial D_4) \nonumber 
\ee 
is closed under $\partial$.
This tells us that any equation $\partial(U_4,U_3) = (W_3,W_2)$
for a {\em fixed} representative of a homology class  
\be 
[(W_3,W_2)] \in H_3(Y,W_7;\ZZ) 
\ee 
is invariant under these shifts. As $\partial T_8
= 0$ we see that we are allowed to modify $V_8$ by
deformations that do not change the boundary. This fixes $V_8$ to
be any representative of a given cobordism class in $Y$, {\it
i.e.~}$\partial V_8 = \text{{\em something fixed}}$. On the other
hand, the equation $\partial (-D_8 + T_8 + Z_8 ) = 0$ states that
the equation $W_7 - V_7 +\partial V_8 = \text{{\em anything
fixed}}$ remains invariant under all deformations. This means that
the cobordism class of $V_8$ is actually given by $W_7-V_7$, and
that $W_7$ and $V_7$ have a common boundary. If we call this boundary
$W_6$, we find $\partial(V_8,W_7)=(V_7,W_6)$.
\end{quote}
This leads us back to (\ref{rel32}) and (\ref{rel76}).

\section{The Symmetry Algebras}
\label{Algebra}

The quadratic tension rule $(T_3)^2 = 2\pi T_6$ can  be
derived also from the M-theory gauge algebra. To see exactly
how, let us look again at the gauge algebra mentioned in Sec.~\ref{Introduction} 
and in Eqs (\ref{galg1}--\ref{galg4}),
extended now into the 12D space $Y$ 
\be
\delta_{\cal G} (C,\newB) &=& (\Lambda_3, \lambda_2 ) \\
\delta_{\cal G} \tC  &=& \Lambda_6 + \half \Lambda_3 \wedge C~.
\ee
The gauge transformations satisfied $\dd \Lambda_3 = \dd \Lambda_6 = 0$  
and $\myiota^*\Lambda_3 = \dd \lambda_2$.    
We have already learned in Eqs (\ref{gtquant1}) and (\ref{gtquant2}) 
that the large gauge transformations live on a lattice
\be
\left[ \frac{T_3}{2\pi} (\Lambda_3, \lambda_2 ) \right] &\in&  H^3(Y, W_6; \ZZ) \\
\left[ \frac{T_6}{2\pi} \Lambda_6 \right] &\in&  H^6(Y,\ZZ) ~. 
\ee 
We distinguish here between a cohomology class, {\it e.g.}~$[\Lambda_6]$, and its
representative $\Lambda_6$.  These gauge transformations satisfy a
{\em noncommutative algebra}, where the only nontrivial bracket is
\be 
[\delta_3',\delta_3] &=& \delta_6 \label{alge} ~,
\ee
with structure constants such that  $\Lambda_6 = \Lambda_3' \wedge
\Lambda_3$. This leads us to the derivation of the tension rule
(\ref{quant2}) as presented in \cite{Lavrinenko:1999xi}: namely that
since $\Lambda_3' \wedge \Lambda_3$ has to belong to the same
lattice as $\Lambda_6$, we get
\be
 \frac{T_3}{2\pi} \chi \wedge \frac{T_3}{2\pi} \chi'
 &\stackrel{!}{=}&    \frac{T_6}{2\pi}   \psi
\ee
for some representatives of the integral elements 
$[\chi],[\chi'] \in H^3(Y,\ZZ)$, $[\psi]
\in H^6(Y,\ZZ)$. This holds if and only if  
\be
T_3^2 &=& 2\pi~T_6 \label{d2} ~.
\ee
For illustration, see Fig.~\ref{fig10}: $\chi$ and $\chi'$ could be  the volume forms of
two independent 3-spheres or 3-tori in the handle $\Sigma_7$,  one of which, $S^3$, is
indicated in the picture.  It is quite remarkable that the quantisation condition
(\ref{quant2}) arises in this way. Our treatment of this derivation adds to that
presented in \cite{Lavrinenko:1999xi} a consistent treatment of global gauge
transformations that was not addressed there.

Though we have not investigated the intricacies of diffeomorphisms 
and anomaly cancellation in this paper, it is of interest to see how the 
gauge symmetry algebra fits together with the diffeomorphism algebra. Using notation from 
App.~\ref{Vols}
the diffeomorphism transformations act on the various fields
according to
\be
\delta_{\cal D} (C,\newB) &=& \frac{T_3}{2\pi}~ \Big(\dd
\Omega_2^{(1)}, ~ 0 \Big) \\
\delta_{\cal D} \tC  &=&  \frac{T_6}{2\pi}~ \Omega_6^{(1)}  +
\half \dd \Omega_2^{(1)} \wedge \frac{T_3}{2\pi}~ C ~.
\ee
These rules follow, as explained above and in App.~\ref{Vols}, 
from requiring that $\tG$
be invariant and that $G$ look like $\kappa T_\sst{6}~
\Omega_\sst{4}$ near the five-brane. 
As $\newB$ is purely worldvolume data, it is natural to keep it 
inert under these diffeomorphisms. 
From the five-brane worldvolume
point of view, these transformations reduce to worldvolume
diffeomorphisms and SO(5) gauge transformations on the normal
bundle as suggested in the pull-back of the bulk tangent bundle
$T_\sst{X}|_{\sst{W}} = T_\sst{W} \oplus NW$. The difference
between large and small diffeomorphisms is that the former
\cite{Bonora:1997kx,Bonora:2000ie} preserve the normal bundle $NW$
and $\Omega_\sst{2}^{\sst{(1)}}$ is globally defined, whereas in
the latter only $\delta \Omega_\sst{3}$ exists globally.

The  transformation functionals are not inert under repeated
diffeomorphism transformations, but instead one finds the non-trivial
contribution
\be
\delta_{\cal D}^{(2)} \tC  &=& \frac{T_6}{2\pi}~ \dd \Big(
\Omega_5^{(2)} + \half \Omega_2^{(1)} \wedge \dd
\Omega_2^{(1)\prime} \Big)  ~.
\ee
We conclude that iterated diffeomorphisms reduce to gauge
transformations $\delta_{\cal D}^{\sst{(2)}} = \delta_{\cal G}$ with
the parameter
\be
\Lambda_6 &=& \frac{T_6}{2\pi}~ \dd \Big( \Omega_5^{(2)} + \half
\Omega_2^{(1)} \wedge \dd \Omega_2^{(1)\prime} \Big) ~.
\ee
Also, diffeomorphisms and gauge transformations do not
commute, but
\be
[\delta_{\cal D},\delta_3] &=& \delta_6 ~,
\ee
where $\Lambda_\sst{6} =  ({T_\sst{3}}/{4\pi})~
\dd\Omega_\sst{2}^{\sst{(1)}} \wedge \Lambda_\sst{3}$. This shows,
therefore, that   the algebra (\ref{alge}) remains valid also for
the direct sum of transformations 
\be
\delta &=& \delta_{\cal D} + \delta_{\cal G}~
\ee
of diffeomorphism and gauge transformations. The real form of this algebra 
is a truncation of the algebras found in 
Refs \cite{Cremmer:1997ct,Cremmer:1998px,Julia:2000af,West:2001as,Damour:2002cu}, 
which include for instance $E_{11}$ and hyperbolic Ka\v{c}--Moody algebras. 
The integral forms 
of these algebras arise in quantum string theory in ten dimensions 
as U-duality.

\section{Discussion}
\label{Discussion}

Let us start by recapitulating our results. 
The correct, though not {manifestly} gauge invariant action is 
\be
S_\gauge &=&  
         ~\frac{1}{\kappa} \int_X  \frac{1}{2}~ G \wedge * G 
       - \frac{1}{6} ~C \wedge G \wedge G
       + \kappa T_3 G \wedge \gravano_7 
       \nonumber \\
  & &  \qquad \qquad \qquad 
       + \frac{T_6}{6} \int_{V_7}  C \wedge  G +
       2\pi       \int_{V_8} \gravano_8 
       \label{total} \\
  & &  + \frac{T_6}{2} \int_{W_6} \left( \half \cH \wedge *\cH 
       - \cH \wedge \myisharp C \right) 
       - T_3 \int_{W_3} C 
       + \frac{T_3}{2} \int_{W_2} \newB  ~. 
       \nonumber
\ee 
In writing the action in this form, it has to be understood that $V_7$ is related 
to $W_7$ and $W_8$ through Eq.~(\ref{rel76}). Using the same relation, the last terms 
of the first and the second line combine to integrals over $V_7$ as well. 
We choose nevertheless to leave an explicit $V_8$ integral here; a purely 
gravitational shift in the action does not change its gauge symmetry properties, 
but will be significant in any future discussion of diffeomorphism symmetry and anomalies. 

Two features in particular merit to be pointed out:  
First, there is no explicit magnetic $\tC$ coupling. 
Secondly, there is a relative factor of two in the 
membrane tension and the worldsheet string tension. 
This latter fact is related to the self-duality of the 
closed worldsheet string theory. 

The above action can be rewritten using essentially only Stokes' theorem 
in the manifestly gauge invariant form
\be
 S_\gauge &=& 
       \int_X \frac{1}{2\kappa}~ G \wedge * G 
     + {T_3} G \wedge \gravano_7(R) 
     - \int_Y \frac{1}{6\kappa} G \wedge G \wedge G  
       \nonumber \\
& &  - \half T_6 \int_{(V_8,W_7)} 
       \Big(  G \wedge G - 2\kappa T_3 ~\gravano_8, 
               ~ \newH \wedge \myiota^* G \Big)    
       \label{final} \\
& &  + \frac{T_6}{4}  \int_{W_6} \newH \wedge *\newH 
     + {T_3}  \int_{U_4} G 
     - \frac{T_3}{2} \int_{U_3} \newH  ~.  
       \nonumber
\ee 
The integration domains here are related to the physical branes by 
\be 
\partial(U_4,U_3) = (W_3,W_2), \quad
\partial(V_8,W_7) = (V_7,W_6), \quad \text{and} \quad \partial Y =X ~. 
\nonumber
\ee 
As was shown in Sec.~\ref{Dependence} this information can actually be recovered from the
action  (\ref{final}) as well. 

We conclude that the original action $S \mod 2\pi$ in Eq.~(\ref{total}) should be, as far
as it is possible to express in terms of standard differential forms, the physically
correct coupling of brane and bulk dynamics. 

More precisely, we have found the following: 
\begin{itemize}
\item[1)]
We have presented a novel globally well-defined form of the
action, collected in Eq.~(\ref{final}) above, thereby elucidating 
how five-branes should
be embedded into the 11D supergravity background.
Though this action involves spurious surfaces, the standard
physical branes emerge as solutions of ``topological" equations of
motion. The way in which these spurious surfaces appear poses
restrictions on any putative 12D theory on
whose boundary M-theory could arise.
\item[2)]
We have also shown how the M-theory gauge algebra remains valid
globally. This leads to the well-known
\cite{Schwarz:1996jq,deAlwis:1996ez,Lavrinenko:1999xi,Julia:2000af}
constraints on the brane tensions
\be
\kappa ~ T_3 T_6 &=& 2\pi \hbar ~,\\
(T_3)^2 &=& 2\pi \hbar ~ T_6 ~.
\ee
This was checked in Eqs (\ref{quant2},\ref{d1},\ref{d2})
independently  using Wu--Yang type arguments, and also by
requiring independence with respect to the positioning of the
Dirac surfaces.
\item[3)]
As a by-product, we derived an Abelian counterpart
(\ref{addV7}) of the Hopf--Wess--Zumino term proposed in
\cite{Intriligator:2000eq}. It arises naturally in requiring gauge invariance.  
\item[4)]
We identified the five-brane charges as integrals of classes in the
relative cohomology of $(X,W_6)$. Similar conclusions were made
already in Refs \cite{Figueroa-O'Farrill:2001kz,Stanciu:2000fz}
where D2 and D0-branes in WZW models were discussed. In trying to
define charges for these Chern--Simons type systems, 
most previous candidates have failed to be either localised, or 
quantised, or gauge invariant
\cite{Bachas:2000ik,Taylor:2000za,Marolf:2000cb}. On the level of fluxes,   
relative cohomology seems to provide the right classification of
physical states and a resolution to these difficulties.
\end{itemize}

\subsection{No new degrees of freedom}

The inclusion of Wess--Zumino terms defined on 
$W_7 \subset X$ might appear surprising, as they do not correspond to
physical M-branes. Adding these geometric objects is nevertheless
perfectly consistent with what we know about M-theory as long as
they do not introduce new degrees of freedom. Their ``worldvolume''
theories, of which we have seen only a part, could perhaps turn
out to be gravitational or topological theories in nature, 
whose dynamical degrees of freedom have a holographic 
description on their boundary, that is to say on the five-brane. 

As these additions are expressible purely as integrals  
over $V_7$, we should now discuss the differences of these 
Dirac surfaces and in particular the r\^ole that this 
boundary $V_7$ plays in more detail.

As we have discovered, the word ``boundary'' may stand for many things in an analysis
involving relative cohomology. The Dirac surface here is actually the relative class
$(V_8,W_7)$ and the physical brane worldvolume is the class $(V_7,W_6)$.  However, all
the modifications that we have  actually made to the standard M-theory action in eleven
dimensions involve $V_7$ rather  than one of the pair $(V_8,W_7)$. Nevertheless, $V_7$ is
just a Dirac surface for the five-brane, but it has an interesting origin: most
conspicuously, perhaps, $V_7$ extends out of the 11D spacetime $X$ of M-theory, and is
therefore inherently an object of a 12D theory. It is physical in the sense that the lift
of the Bianchi identity (\ref{Ceom}) to twelve dimensions, as given in (\ref{defV7}),
involves a delta-function onto $V_7$, so that the five-brane worldvolume is essentially
just the dimensional reduction of this to eleven dimensions. Let us discuss next what
this might mean.

There is a final superalgebra in twelve dimensions 
\cite{vanHolten:1982mx} that still has 32-component spinorial generators. 
This 12D superalgebra differs crucially from the 
11D one in that the signature of the metric is $10+2$ rather than
Minkowskian and that it does not contain Poincar\'{e} generators.
The absence of Poincar\'{e} generators in the rigid
12D algebra is suggestive of a topological theory 
with $\SO(10,2)$ invariance. 
This may suggest that the theories in twelve and eleven
dimensions could be related by some form of Holography, which
would preemptively resolve the apparent difficulties with the appearance of new
degrees of freedom.

Furthermore, the 12D superalgebra contains in particular a self-dual 6-form
central charge. It is therefore conceivable that 
$V_7$ could appear as a BPS state in that 
theory.\footnote{Because of the signature, the treatment of BPS
states would need a more thorough analysis, so here we restrict
the discussion just to dimension counting. In
Ref.~\cite{Hewson:1997wv} it was argued that the BPS states would
be F3- and F7-branes instead; perhaps we should
think of $U_4$ and $V_8$ as the Dirac surfaces that make contact
with the (yet unknown) dynamics in twelve dimensions.}  
This would be natural as the 6-form central charge reduces exactly to a five-brane 
charge under (double) dimensional reduction. This reduction does not  
give rise to an 11D 6-form central charge precisely because of 
the self-duality condition: we can always choose to work in a basis where 
the 12D 6-form involves the twelfth, compact coordinate. 
This emphasises further the fact that the 12D formulation does not 
import new degrees of freedom to the theory, since even on the level of representations
of  the respective superalgebras the 5-form and the 6-form central charges describe 
the same number of dynamical degrees of freedom.

\subsection{Open questions} 

In dealing with smooth manifolds with boundaries, the boundary operator is nilpotent
$\partial^2=0$.  Nevertheless, much of the interesting structure of 11D formulations of
M-theory is related to the  Ho\v{r}ava--Witten boundaries of $X$, which we have not
addressed here at all.  They could be included in the discussion replacing, in principle,
smooth manifolds with manifolds with edges and vertices. This would take us into the
realm of extended topological quantum field theory,  possibly along the lines of
Ref.~\cite{kerler}. An early indication  of this is perhaps the fact  that we could think
of the five-brane worldvolume as the edge in the manifold $W_7-V_7$ sketched in
Fig.~\ref{fig1},  thus avoiding the need to  assume that the ``Dirac surface for Dirac
surfaces'' $V_8$ exist. 

Diffeomorphism symmetry and its violations by five-brane solutions play a pivotal
r\^{o}le in anomaly cancellation. Analysing this requires understanding fully the
behaviour  of the $C$-field solution near a five-brane worldvolume \cite{Freed:1998tg}. 
In addressing these issues, Wess--Zumino terms built out of worldvolume scalars are
crucial \cite{Hassan:2000zk}.  Another interesting question is when exactly are the
extensions of worldvolume fields to Dirac surfaces obstructed. In the case of the
worldvolume field $\newH$ this issue can be settled using Hodge decomposition,  but it
becomes more subtle in the case of the worldvolume scalars. 

A formulation of the theory in terms of non-dynamical Dirac surfaces may 
involve higher dimensional physical states, such as BPS states, as long as these states  
reduce to the same number of physical states as that expected in 
the original lower dimensional formulation of the theory. 
The problem is then to find a mechanism to guarantee that this always happens. 
As was pointed out in Sec.~\ref{hwz} and in \cite{Intriligator:2000eq}, 
the term $\int C\wedge G$ on a classical solution calculates the winding number of 
$V_7$ on $S^4$. An obvious candidate for the holographic dual of
the five-brane worldvolume theory on $V_7$ could therefore be a
topological quantum field theory based on the group $\pi_7(S^4)=\ZZ\oplus\ZZ_{12}$.
The idea of formulating M-theory as a holographical theory has
surfaced earlier in a different context, for instance in \cite{Horava:1997dd}.

\vspace{5mm}
\noindent {\bf Acknowledgements:}  

We thank T.~Damour, B.~Julia and R.~Minasian for helpful discussions. We also wish to
acknowledge the hospitality  of the Institut des Hautes \'Etudes Scientifiques  and of
the Isaac Newton Institute at times during the course of this work.  J.K.~acknowledges
the financial  support provided through the European Community's Human Potential
Programme under contract HPRN-CT-2000-00131 Quantum Spacetime.

\pagebreak
\appendix

\section{Appendices}

\subsection{Currents and delta-functions}
\label{Currents}

Given a subsurface $\Sigma$ without boundary, {\it i.e.~}a cycle
in the homology of a surface $M$, {\em Poincar\'e duality}
produces a differential form $\delta_M(\Sigma)$ in the (compact)
cohomology on the surface, such that all integrals over $\Sigma$
can be converted into integrals over $M$
\be
\int_\Sigma \omega &=& \int_M \omega \wedge \delta_M(\Sigma)~.
\ee
As we are dealing with subsurfaces with boundaries, we need a
more general construction.

Representatives of the ``delta-functions'' $\delta(W_6)$ can be explicitly written down
given the embedding of the brane worldvolume $W_6$ into the spacetime in
parallel, or physical, gauge $X(\sigma)^i=\sigma^i$ for $i=0,\ldots,5$. Then
$X^\mu$ for $\mu=6,\ldots,10$ are the normal coordinates and we have 
\be
\delta(W_6) &=&  \delta({X^6}) \dd X^6 \wedge \cdots \wedge
\delta({X^{10}}) \dd X^{10} ~.
\ee
The $\delta$'s on the left are now the standard Dirac
delta-functions in one dimension; the notation on
the left indicates that the whole object is a differential form with rank
equalling the codimension of the brane; it can therefore be
equated to Bianchi identities in form-field notation.

In a similar vein we can also describe surfaces with boundaries: if
the five-brane worldvolume is a boundary of a seven-surface $W_7$ located at
$X^{6}=0$, we can describe it using
\be
\delta(W_7) &=&  -\vartheta({X^6}) ~ \delta(X^7) \dd X^7 \wedge
\cdots \wedge \delta({X^{10}}) \dd X^{10}\label{deltaw7}  ~, 
\ee
where the Heaviside theta-function $\vartheta$ equals 1 on
non-negative arguments and vanishes otherwise.  This also ensures
that the general formula $\dd \delta(W_p) = (-1)^{p}
\delta({\partial W_p})$ applies, as required by Stokes'
theorem.

The situation is slightly more complicated if the spacetime itself 
has a boundary. In particular, considering $\partial Y=X$ we see again
by Stokes' formula
\be
\delta_Y(\partial W_p) &=& (-)^{p} \dd \delta_Y(W_p) +
\delta_Y(W_p) \wedge \delta_Y(X) \label{fullStokes} ~. 
\ee

In studying the various Dirac surfaces we need to lift
delta-functions from $X$ to $Y$. This involves the inclusion
\be
\myiota : X \hookrightarrow Y ~.
\ee
Given that $\partial W_7 = W_6$ and $W_7 \subset X$, we can always
find similarly a subsurface $V_7$ such that $\partial V_7 = X
\cap V_7 = W_6$ and $V_7 \subset Y$; then one has
\be
\delta_X(W_6) &=& - \dd ~ \delta_X(W_7) \\
 &=& + \myiota^*  \delta_Y(V_7)\ .
\ee
It follows then that
\be
\delta_X(W_7) &=& + \myiota^*  \delta_Y(V_8) 
\ee
and that the boundary of $V_8$ can be reconstructed by gluing $W_7$
and $V_7$ together along their common boundary $W_6$, with the
orientation of $V_7$ inverted: we denote this operation by
\be
W_7 - V_7 &=& \partial V_8 \label{defV8}\ .
\ee
This leads us to the formalism of relative cohomology, {\it
cf.~}App.~\ref{Relative}; we observe that all of this can be
summarised in the statement
\be
\partial (V_8,W_7) &=& (V_7,W_6)\ .
\ee

In the text we have freely calculated sums of subsurfaces
multiplied with real numbers as is usual in real homology. If
$\alpha,\beta \in \RR$, expressions like $\alpha W_3 +
\beta U_3$ occur in integrals or delta functions and refer to
operations such as
\be
\int_{\alpha W_3 + \beta U_3} \omega &\equiv& \alpha \int_{W_3}
\omega + \beta \int_{U_3} \omega ~. \label{c-on-n}
\ee

\subsection{Chern--Simons functionals}
\label{CSfnals}

Chern--Simons functionals on odd-dimensional surfaces $M$ assign
a phase in $\RR/2\pi \ZZ$ to gauge fields. Let us considering an
Abelian $n$-form gauge field $A$ for instance: given a bounding
surface $N$ such that $M=\partial N$ we can look at integrals of
invariant polynomials of the curvature $F=\dd A$, such as
\be
\frac{2\pi}{n!} \int_N F^{\wedge n} \label{instanton}\ .
\ee
Locally these functionals can be put in the form
\be
\frac{2\pi}{n!} \int_M A \wedge F^{\wedge n-1} \label{naive}
\ee
using the Bianchi identity $\dd F =0$. The ambiguity in the
choice of the surface $N$ under redefinitions differing by 
addition of a closed surface gives rise to a quantisation
condition
\be
F &\in& H^*(X, \ZZ) ~. 
\ee

In this article, we have been particularly interested in situations
where a connection $A$ with curvature $F=\dd A$ is well-defined
only outside a fixed singular locus $W$. In particular, we
assume
\be
\dd F &=& \delta(W) \in H^*_c(M)
\ee
in the compact cohomology of the spacetime surface $M$. We can
still define Chern--Simons functionals by gauge invariant
expressions of the type (\ref{instanton}), even though they can no
more be put into the form (\ref{naive}). 
To do this, we would need to modify
the theory by adding source terms on the singular locus $W$.

We may nevertheless continue to consider the traditional
Chern--Simons terms on a specific class of surfaces $M-W$ that allow a
bounding surface $N$ such that $\delta_W$ has no support on it.
In this case the quantisation condition is
\be
F &\in& H^*(X-W, \ZZ)\ .
\ee

A more formal way of studying this ``punctured" cohomology
$H^*(X-W)$ is actually in terms of the {\em relative cohomology}
$H^*(X,W)$, {\it cf.~}App.~\ref{Relative}. This is also where
field theory considerations naturally lead. If we thicken $W$ to a
small tubular neighbourhood of the actual brane worldvolume, the excision
axiom of homology states that this homology theory is insensitive
to the structure of the brane worldvolume $W$ itself: it only sees the
surface with the brane worldvolume removed from it, with a small tubular
neighbourhood, essentially $H_*(X-W)$.\footnote{To put this in a
more precise manner, if the tubular neighbourhood is a disc bundle
$D^{n+1}W\longrightarrow W$, and its boundary defines a sphere
bundle $S^nW\longrightarrow W$, then it is true that
$H_*(X,\closure({D^{n+1}W})) \simeq H_*(X-D^{n+1},S^nW)$. This is
not quite the cohomology of $X-W$ or $X-D^{n+1}W$, but rather the
cohomology of the relative pair of the surface minus the tubular
neighbourhood $X-D^{n+1}$ and the boundary of the tubular
neighbourhood $S^nW$. This essentially amounts to removing the
surface $W$ from the spacetime $X$.}

Let us consider, more concretely, the case where $W$ is a brane worldvolume of
codimension $n+1$. Then, a nontrivial $F$ can only be an integer
multiple of the volume form of the transverse sphere $S^n$, {\it
i.e.}, the generator of $H^{n}(X-W, \ZZ)$. This means that the
gauge field $A$ cannot be well-defined on $W$. We may decompose it
locally into singular and the smooth parts
\be
C =  \Omega_{n-1} + \hat C ~,
\ee
where $\dd \Omega_{n-1} = \Omega_n = \Vol(S^n)$ and $\hat C$ is
well-defined near $W$. It turns out that $\Omega_n \sim * \dd
\rho^{n-1}$, where $\rho$ is the transverse distance from $W$ and
$*$ is the Hodge star in the transverse space. For the standard
Dirac monopole, we have $F= \dd \cos\vartheta \wedge \dd\varphi /4
\pi$ and $A = \cos\vartheta \wedge \dd\varphi /4 \pi$, for
instance. The coordinate system $\vartheta,\varphi$ is not defined
at the origin, $W = \{\vec{0}\} \in \RR^3$, where the monopole
lies.

These ideas can be formalised  in terms of Chern kernels \cite{HL}
to some extent. We can think of the brane-charge of $W$ then as
the {\em residue} of the curvature of the singular connection $C$.

\subsection{Volume forms}
\label{Vols}

Any closed differential form $\Omega_{\sst n}$, $\dd \Omega_{\sst n-1} =0$ can be locally 
expressed in terms of a potential $\Omega_{\sst n} =\dd \Omega_{\sst n-1}$. 
Any change $\delta \Omega_{\sst n-1}$ in the choice of the representative of the local potential 
must be closed $\delta \Omega_{\sst n-1} = \dd \Omega_{\sst n-2}^{{\sst (1)}}$. 
More generally, repeated changes of representatives (gauge transformations) give rise to a ladder 
$\delta \Omega_{\sst n-k} = \dd \Omega_{\sst n-k-1}^{{\sst (k)}}$. 

Let us consider the case of an isolated, though not necessarily trivially embedded
five-brane. We can always parametrise its worldvolume by local coordinates 
$\sigma^i, i=0,\ldots,5$; close to the brane worldvolume there also always exists a good
system of local transverse coordinates $y^a$ \cite{LM}. If the brane worldvolume is
nontrivially embedded,  its normal bundle will have curvature. This can be expressed by
giving a covariant connection $D_A$  with field strength  $F_A$. It acts on sections
$\phi^a$ of the bundle as 
\be
D_A \phi^a(\sigma) &\equiv& \dd \sigma^i \left[ \partial_i ~\phi^a(\sigma)  + 
A_i(\sigma)^{ab}  \phi^b(\sigma) \right] ~. \label{covder} 
\ee
The volume form of the transverse four-sphere $\Omega_{\sst 4}$ can now be expressed in the local coordinate system 
$(\sigma^i, y^a)$ as  \cite{Freed:1998tg}
\be 
\Omega_4 &=& \frac{1}{64 \pi^2}
\epsilon_{a_1a_2a_3a_4a_5} \Big( D_A \hat{y}^{a_1}  \wedge D_A
\hat{y}^{a_2} \wedge D_A  \hat{y}^{a_3} \wedge D_A  \hat{y}^{a_4} \nonumber\\
  & &
+ 2 F^{ a_1a_2}_A \wedge
 D_A  \hat{y}^{a_3} \wedge D_A \hat{y}^{a_4} +
 F^{a_1a_2}_A \wedge F^{a_3a_4}_A
 \Big) ~\hat{y}^{a_5}  \label{omega4} ~,
\ee
where $\hat{y}^a = y^a |y|^{-1}$ is the unit normal
vector. It has a local potential $\Omega_{\sst 4} =\dd \Omega_{\sst 3}$ 
which can be expressed as \cite{Freed:1998tg}
\be
\Omega_3 &=& \frac{1}{32 \pi^2}
\epsilon_{a_1a_2a_3a_4a_5} \Big(  A^{a_1a_2}  \wedge \dd
A^{a_3a_4} ~ \hat{y}^{a_5} - \half A^{a_1a_2}  \wedge
A^{a_3a_4}
 \wedge \dd \hat{y}^{a_5} \nonumber \\
 & & + 2 A^{ a_1a_2}  \wedge
 \dd  \hat{y}^{a_3} \wedge \dd \hat{y}^{a_4}  ~ \hat{y}^{a_5}
 \Big) \label{omega3} ~.
\ee 
In diffeomorphisms of $X$ this local potential changes by a locally exact term 
$\delta\Omega_{\sst 3} =\dd \Omega_{\sst 2}^{{\sst (1)}}$ 
We quote again Ref.~\cite{Freed:1998tg} for the explicit form of
$\Omega_\sst{2}^{\sst{(1)}}$
\be
\Omega_2^{(1)} &=& \frac{1}{16\pi^2}
\epsilon_{a_1a_2a_3a_4a_5} \varepsilon^{a_1a_2} \Big(
 \dd  \hat{y}^{a_3} \wedge \dd \hat{y}^{a_4}  ~ \hat{y}^{a_5}
+     A^{a_3a_4}
 \wedge \dd \hat{y}^{a_5}
 \Big) \label{defO2} ~,
\ee
where $\varepsilon^{\sst{a_1a_2}}$ is the $\so(5)$ gauge
transformation parameter, $\delta A = D_\sst{A} \varepsilon$.

The volume form $\Omega_{\sst 4}$ is obviously 
ill-defined on the brane worldvolume. Nevertheless, it is sufficient 
to the purposes of the present paper to choose the formal convention 
that
\be
\myiota^*\Omega_4 &=& 0 ~. 
\ee 
After all, the form $\Omega_{\sst 4}$ is proportional to differentials 
of transverse coordinates. This simplification means that we will have to postpone the
discussion of diffeomorphisms and anomaly cancellation to further work.

\subsection{Relative cohomology}
\label{Relative}

We have organised the structure of the theory naturally in terms
of the relative (co)homology of the 11D surface
$X$ with respect to the five-brane's worldvolume $W_6$. This means that the
natural objects to consider in homology are not, necessarily,
closed subsurfaces of $X$, but rather such surfaces $\newCC_k \subset
X$ that may have a boundary inside the five-brane worldvolume $\partial \newCC_k =
\newC_{k-1} \subset W_6$. As an integrability condition one has to
impose $\partial \newC_{k-1} = 0$. Such pairs $(\newCC_k,\newC_{k-1})$ are
representatives of a class in the relative homology\footnote{For
an introduction to relative (co)homology we refer to  textbooks on
topology such us Refs \cite{Bredon,Bott}. For an application to  
string theory see \cite{Figueroa-O'Farrill:2001kz}.} group
$H_k(X,W_6;\ZZ)$. The boundary operator acts as
\be
\partial(\newCC_k,\newC_{k-1}) &\equiv& (\newC_{k-1} - \partial \newCC_{k}, \partial \newC_{k-1})
~.
\ee
More geometrically this means that we simply choose to neglect
whatever topology the five-brane worldvolume might support, as if
contracting the five-brane worldvolume to a point.

The dual of this construction is the relative cohomology of $X$
with respect to $W_6$: there we consider pairs of forms
$(\Omega_k, \omega_{k-1}) \in H^k(X,W_6; \ZZ)$ that satisfy 
\be
\dd \Omega_k &=& 0 \\
\myiota^*\Omega_k - \dd \omega_{k-1} &=& 0 ~,
\ee
where $\myiota^*$ is the pull-back by the inclusion $\myiota : W_6
\hookrightarrow X$. This means that the closed forms that we see in the
bulk should be cohomologically trivial when pulled back onto the
five-brane worldvolume. The nilpotent coboundary operator of this cohomology
acts as
\be
\dd (\Omega_k, \omega_{k-1}) &\equiv& (\dd \Omega_k, \myiota^*\omega_k
- \dd \omega_{k-1}) ~.
\ee

Finally, there is a natural pairing of homology with cohomology:
\be
\int_{(\newCC_k,\newC_{k-1})} ~ (\Omega_k, \omega_{k-1}) &=& \int_{\newCC_k}
\Omega_k -\int_{\newC_{k-1}} \omega_{k-1}\ .
\ee
This paring is naturally invariant under a change in the
representative of the (co)homology class, as can be easily seen by
shifting either one of the classes by an exact term. Stokes'
formula acquires an extra sign with these conventions.

The long exact sequence of relative cohomology
\be
\stackrel{\myiota^*}{\longrightarrow} H^{k-1}(W_6)
\stackrel{j^*}{\longrightarrow} H^{k}(X,W_6)
\stackrel{p^*}{\longrightarrow} H^{k}(X)
\stackrel{\myiota^*}{\longrightarrow} H^{k}(W_6)
\stackrel{j^*}{\longrightarrow}\ , \label{cles}
\ee
where the mappings are the coboundary map $j^*\omega_{k-1} =
(0,\omega_{k-1})$, the obvious projection $p^*(\Omega_k,
\omega_{k-1})=\Omega_k$ and $\myiota : W_6 \hookrightarrow X$, guarantees
that we can naturally transport terms from one cohomology group
to another, and that their integrality properties are thus preserved. This
will be important in finding relations between charge lattices.
There is a similar long exact sequence in homology
\be
\stackrel{j_*}{\longrightarrow} H_{k}(W_6)
\stackrel{\myiota_*}{\longrightarrow} H_{k}(X)
\stackrel{p_*}{\longrightarrow} H_{k}(X,W_6)
\stackrel{j_*}{\longrightarrow} H_{k-1}(W_6)
\stackrel{\myiota_*}{\longrightarrow}~, \label{hles}
\ee
with $\myiota_*$ the inclusion, $p_* \newCC_k = (\newCC_k,0)$, and $j_*(\newCC_k,\newC_{k-1})
= \newC_{k-1}$.

\subsection{Gravitational corrections}
\label{Gravitational}

The 11D supergravity action \cite{Cremmer:1978km}
gets higher derivative corrections. Of these, the one relevant to our present discussion is 
\be
S_\corr &=&  T_3 \int_X  C \wedge \gravano_8(R)\ . \nonumber
\ee
This correction has been found from  T-duality considerations in
Ref.~\cite{Duff:1995wd}, and is given by 
\be
{\gravano_8(R)} &=& \frac{1}{4! ~8} \Big(p_1^2 - 4 p_2\Big)
\label{omega8} \\
&=& 
 \frac{1}{192 ~(2\pi)^4} \Big(\tr R^4  - \frac{1}{4} (\tr R^2)^2  \Big)
\ . \nonumber 
\ee 
The Riemann curvature $R$ is seen here as an $\so(8)$-valued 2-form.

This correction can straightforwardly be incorporated in the 
analysis of this paper by redefining the magnetic field strength as
\be
\tG &\equiv& \dd \tC + \half C \wedge G - \kappa T_3 \gravano_7 ~,
\ee
using $\gravano_8 = \dd \gravano_7$. This correction will give rise
to membrane tadpoles \cite{Witten:1997md}, which have to be
cancelled in the absence of other sources. Therefore,
consistent backgrounds must satisfy
\be
\frac{T_3}{48}(\lambda^2-p_2) \in H^8(X,\ZZ) ~,
\ee
using $\lambda =p_1/2$ as in Ref.~\cite{Witten:1997md}. If we evaluate
this on an eight-surface $M \subset X$ that supports nowhere-vanishing
spinors then \cite{Isham:1988qe}
\be
 \frac{1}{48} \int_M \lambda^2-p_2 &=& \mp \frac{\chi(M)}{24} ~,
\ee
where $\chi$ is the Euler characteristic. This  depends on the
representation of the nowhere-vanishing spinor: the minus
(resp.~plus) sign corresponds to the assumption that the nowhere-vanishing spinor
be in the $\mathbf{8}_s$ (resp.~$\mathbf{8}_c$) representation of Spin(8).

There is also a shifted integrality condition
\cite{Witten:1997md}
\be
\left[T_3 \frac{G}{2\pi}\right] - \frac{\lambda}{2} \in H^4(X,\ZZ)
\label{lshift}
\ee
in the absence of five-brane sources. This follows from
considerations involving $E_8$ bundles on Ho\v{r}ava--Witten
boundaries \cite{Horava:1996ma,Horava:1996qa}.

\end{document}